\documentclass[table]{article}
\usepackage[a4paper, margin=1in]{geometry}
\usepackage{authblk}
\usepackage{booktabs}
\usepackage{amsmath}
\usepackage{amssymb}
\usepackage{pgfplots}  
\usepackage{todonotes}
\usepackage[hyphens]{url}
\PassOptionsToPackage{table}{xcolor}
\pgfplotsset{compat=1.9}
\usepackage{caption}
\usepackage{subcaption}
\providecommand{\keywords}[1]{\textbf{\textit{Index terms---}} #1}
\usepackage[english]{babel}
\usepackage{graphicx}
\usepackage{dirtytalk}
\usepackage{listings}
\usepackage{xcolor}

\colorlet{punct}{red!60!black}
\definecolor{background}{HTML}{EEEEEE}
\definecolor{delim}{RGB}{20,105,176}
\colorlet{numb}{magenta!60!black}

\lstdefinelanguage{json}{
    basicstyle=\normalfont\ttfamily,
    showstringspaces=false,
    breaklines=true,
    backgroundcolor=\color{background},
    literate=
     *{0}{{{\color{numb}0}}}{1}
      {1}{{{\color{numb}1}}}{1}
      {2}{{{\color{numb}2}}}{1}
      {3}{{{\color{numb}3}}}{1}
      {4}{{{\color{numb}4}}}{1}
      {5}{{{\color{numb}5}}}{1}
      {6}{{{\color{numb}6}}}{1}
      {7}{{{\color{numb}7}}}{1}
      {8}{{{\color{numb}8}}}{1}
      {9}{{{\color{numb}9}}}{1}
      {:}{{{\color{punct}{:}}}}{1}
      {,}{{{\color{punct}{,}}}}{1}
      {\{}{{{\color{delim}{\{}}}}{1}
      {\}}{{{\color{delim}{\}}}}}{1}
      {[}{{{\color{delim}{[}}}}{1}
      {]}{{{\color{delim}{]}}}}{1},
}
\usepackage{amsmath}

\usepackage{booktabs}
\usepackage{subcaption}
\usepackage[hyphens]{url}
\usepackage{todonotes}
\usepackage{floatrow}
\usepackage{hyperref}

\title{The Anatomy of Deception: Technical and Human Perspectives on a Large-scale Phishing Campaign}

\author[1]{Anargyros Chrysanthou}
\author[2]{Yorgos Pantis}
\author[1,2]{Constantinos Patsakis}
\affil[1]{Department of Informatics, University of Piraeus, 80 Karaoli \& Dimitriou str., 18534 Piraeus, Greece}
\affil[2]{Information Management Systems Institute of Athena Research Centre, Greece}

\usepackage{pgfplots, pgfplotstable}
\definecolor{col1}{HTML}{ef476f}
\definecolor{col2}{HTML}{118ab2}
\definecolor{col3}{HTML}{ffd166}
\definecolor{col4}{HTML}{06d6a0}
\definecolor{col5}{HTML}{073b4c}

\pgfplotsset{compat=1.18}
\begin{document}
\date{}
\maketitle
\begin{abstract}
In an era dominated by digital interactions, phishing campaigns have evolved to exploit not just technological vulnerabilities but also human traits. This study takes an unprecedented deep dive into large-scale phishing campaigns aimed at Meta's users, offering a dual perspective on the technical mechanics and human elements involved. Analysing data from over 25,000 victims worldwide, we highlight the nuances of these campaigns, from the intricate techniques deployed by the attackers to the sentiments and behaviours of those who were targeted. Unlike prior research conducted in controlled environments, this investigation capitalises on the vast, diverse, and genuine data extracted directly from active phishing campaigns, allowing for a more holistic understanding of the drivers, facilitators, and human factors. Through the application of advanced computational techniques, including natural language processing and machine learning, this work unveils critical insights into the psyche of victims and the evolving tactics of modern phishers. Our analysis illustrates very poor password selection choices from the victims but also persistence in the revictimisation of a significant part of the users. Finally, we reveal many correlations regarding demographics, timing, sentiment, emotion, and tone of the victims' responses.   
\end{abstract}

\keywords{
Phishing,  Digital forensics,  Sentiment analysis,  Human factors in cybersecurity}

\section{Introduction}
In the ever-evolving cybersecurity landscape, phishing remains one of the most insidious and prevalent threats. The tactics of cybercriminals have followed technological advancements, leading to a significant escalation in the sophistication of phishing campaigns. These campaigns, which primarily prey on human vulnerabilities, have transcended beyond mere email attacks and now manifest in various forms, ranging from vishing (voice phishing) to more intricate spear phishing, targeting specific individuals or organisations. The widespread implications of these attacks have not only led to significant financial losses, but also undermined trust in digital communication, a cornerstone of modern society. In the academic realm, understanding the mechanics, psychology, and countermeasures associated with phishing is paramount.

This work delves deep into the anatomy of a large-scale phishing campaign targeting Meta's users, but also provides a good insight into the human aspects of this campaign. More precisely, we study the behaviour of victims who responded to a phishing email notifying them that they violated Meta's policies and their account would be terminated. The only way to rectify this would be to complete an appeal form and provide personal data along with their credentials. While there are many studies on the topic, to the best of our knowledge, this is the first to perform such an analysis in a real-world setting and of such scale. To this end, we analyse the technical part of the campaigns, the victims' data, passwords, timing, and demographics. Moreover, since the appeal forms contain a lot of text for further analysis, we try to assess the victims' sentiments and emotions to understand the victims better and determine why they responded. To this end, we apply natural language processing, machine learning, and transformer-based methods to extract insightful information.

As highlighted, the key difference of this work from the existing literature is that the data are from a real, big, and broad set of campaigns. As much as researchers want to replicate a malicious setting, this cannot be achieved as it would lack the scale, diversity, openness, and maliciousness of actual phishing campaigns or put users in a sterile environment where they know that they are somehow being monitored. Indeed, such research is conducted through participatory web or in-person studies or active phishing awareness campaigns. Evidently, all the above introduce various biases and limitations that cannot be easily ignored. On the contrary, our work alleviates most of these constraints as we investigate real targeted phishing campaigns involving more than 25000 Meta's users worldwide. The operational and technical deficiencies of the campaigns enabled us to collect the data that phishers extracted from their victims. Further to merely discussing the campaigns, we attempt to dive into the psychological aspects of the victims by providing a sentiment analysis based on their free text input to the campaigns. Therefore, while we cannot collect fine-grained demographic information about the victims, we can directly analyse their input and actions at scale, drawing clear conclusions about their activity in the phishing campaigns.

As a result, our analysis highlights issues in all layers. First, it illustrates very poor password selection choices from the victims. Beyond the use of easy-to-guess passwords, almost six out of ten passwords have been published in password leaks at least two years ago. Moreover, we observed persistence in the re-victimisation of a significant number of users, as many of them repeatedly added sensitive information to the same phishing site or responded to another email of the set of phishing campaigns. Finally, delving into the victims' input, our analysis reveals many correlations regarding demographics, timing, sentiment, emotion, and tone of the victims' responses, providing fruitful insight into the human aspect of such attacks.  

The remainder of this work is structured as follows. In the next section, we present the current state of the art in phishing attacks, discussing regional statistics and how they match global trends, and then we provide an overview of sentiment and emotion analysis. Next, we describe the phishing campaigns in focus, describing the unique characteristics that led to their success. In Section \ref{sec:vic_stats}, we analyse the victims' input, providing a thorough analysis of the password usage, the timing of the responses, and correlations in terms of demographics. Moreover, we analyse the victims' persistence in interacting with the phishing campaigns. Afterwards, we analyse the text input that the victims provided, to extract the sentiment, emotion, and tone of their text to better understand their state when they were phished. Then, in Section \ref{sec:ops}, we discuss operational issues from the phishers' side and the insights they provide. Moreover, the article concludes by summarising our contributions. Finally, we discuss the lessons learned from the technical and human perspective.  

\section{Related work}
\subsection{Phishing attacks}
Different definitions of \say{phishing} have been proposed and discussed by experts, researchers, and cyber-security organizations. Due to the term's constant evolution, no single definition exists; the term has been multiply defined based on its context and use \cite{alkhalil2021phishing}. Typically, phishing attacks are characterised by manipulating recipients into unwittingly carrying out actions desired by the attackers, referred to as phishers. Phishers use two major strategies to carry out their attacks: psychological manipulation of individuals to extract personal information (commonly known as social engineering) and the utilisation of technical methods to present the intended victims/recipients with an as plausible ploy as possible (a highly believable phishing site, namely a site that adheres to the phishing campaign's content, resembles the mimicked entity; e.g., mimicking the real site of a social network, bank, or other well-known entity). The influence of personal and environmental factors, as well as timeliness, can make some people more susceptible to these attacks, making existing safeguards against phishing frequently fail. Research has been conducted to uncover why this happens and has concluded that human nature plays a big role in phishing, maybe the most important. Beyond time factors and emotional state (e.g. stress \cite{lininger2005phishing}), there is a lot of debate regarding the factors that make individuals more prone to phishing attacks, including but not limited to demographic variables (e.g., age \cite{downs2006decision,sheng2010falls,iuga2016baiting}, gender \cite{sheng2010falls,DBLP:conf/hicss/LiLPGYL20,diaz2020phishing}), and personality characteristics \cite{10142078}. According to many works \cite{iuga2016baiting, ovelgonne2017understanding}, certain personality features can make people more susceptible to different kinds of deceitful strategies. For instance, human beings' susceptibility to greed is a highly exploitable trait \cite{workman2007gaining}. Attackers often take advantage of this by sending emails containing tempting offers, substantial discounts, or free gift cards. Phishing attacks encompass a wide range of types and techniques, including, but not limited to, phishing email and URL obfuscation attacks. All phishing attacks attempt to steal confidential information, simple or even sensitive personal data, and/or credentials, such as financial or social media login information, which the phishers can exploit for their nefarious/fraudulent activities. Email phishing is the most popular type of phishing, as most phishing attacks start with an email \cite{apwg_report} sent to an unwitting victim. The email appears to originate from a trusted source to bypass possible email protection filters. An indicative phishing email, such as the ones studied in this work, could inform the recipient that the latter has violated social media terms and conditions and needs to change credentials.  In most of these emails, the domain name is similar to the real one. For example, phishers may utilise the domain \texttt{goog1e.com}, as the email sender's domain, instead of Google's actual domain, namely \texttt{google.com}, effectively replacing the letter \say{l} with the number \say{one}). This type of change is called typo-squatting and might be missed by the human eye. Similarly, a phishing link might be present in the email, which leads the victim to a new website that again resides in a domain similar to the actual legitimate one and is controlled by the phishing operator \cite{liu2016content, agrawal2016origin}. URL obfuscation attacks are also widely used by modern phishers. Phishers trick victims into clicking on a dubious link that directs them to a malicious phishing server instead of the desired location. In other frequently seen cases, an employee might receive an email containing an alleged recruitment plan in the form of a trojanised Microsoft Office document \cite{krombholz2015advanced,koutsokostas2022invoice}, which will infect the host once it is opened. 

\subsection{Phishing trends in 2022}
To perform attribution in the context of information security, it is crucial to start by observing trends in the activities of malicious actors. A set of criteria essential to the malicious actor's method of operation can then be used. These criteria cover a wide range of variables, including the infrastructure in use, as well as the utilised attack strategies. These factors can be examined to identify particular threat actor groups and start monitoring the actors' operations.

Our research, despite its local nature, was acknowledged by respected peers, such as Volker Weber\footnote{\url{https://vowe.net/2023/03/06/2022-phishing-insights/}} and Brian Krebs\footnote{\url{https://infosec.exchange/@briankrebs/109976564720874444}}, as it tried to look beyond regional interests and differences in order to get a glimpse of the greater landscape. First, it revealed a concerning increase in parcel delivery scams in 2022\footnote{\url{https://v4ensics.gr/phishing/}}, namely in scams where the unwitting victim is informed that an alleged package has arrived but could not be delivered due to missing delivery details. Hence, it is kept in the post warehouses with a relevant fee being owed. The victim is requested to provide his personal details, along with a credit card, to receive the alleged package. The scam spanned across at least 8 European countries (Greece, Hungary, Romania, Slovenia, Finland etc.) and abused the same reputable hosting provider, also abused by the phishing campaign analysed in the present work, namely Google Firebase. 
Information security companies (e.g., ESET\footnote{\url{https://www.welivesecurity.com/2022/10/26/parcel-delivery-scams-know-what-watch-out-for/}}\footnote{\url{https://web-assets.esetstatic.com/wls/2022/10/eset_threat_report_t22022.pdf}}) confirm our research findings, as they report that parcel delivery scams worldwide are increasing, with the most targeted brands being USPS and DHL. The reason is that COVID and the post-COVID-era have boosted e-commerce sales (e.g., there was a 56\% increase between 2019 and 2021), resulting in a surge in package delivery. Fraudsters seized the opportunity and try to trick victims by masquerading as parcel delivery companies. Similarly, police authorities, such as Singapore police\footnote{\url{https://www.todayonline.com/singapore/police-warn-shoppers-about-parcel-delivery-phishing-scams-after-130-victims-lose-s182000-first-2-weeks-december-2072416}} \footnote{\url{https://www.police.gov.sg/media-room/news/20220611_police_advisory_on_phishing_scams_involving_parcel_delivery}} issue announcements/warnings that parcel delivery scams is becoming a trend among scammers, thus citizens need to protect themselves. From January 2022 to June 2022, based on Singapore police, at least 415 victims have fallen prey to parcel delivery scams, with losses amounting to at least \$574,000 US.

Our research showed also that Meta, which includes websites such as Instagram and Facebook, was found to be, global-wise, a popular target for scams. To trick unwary individuals into handing over their Meta account credentials, these attacks featured the development of scam websites that mimicked Meta's services, e.g., Facebook's interface. 
For instance, phishers employed social engineering tactics and notified potential victims that unusual activity was observed on their Instagram account. Thus, they should secure their account by visiting a Meta-like site that the phishers controlled, resembled Instagram's actual page and was utilized by phishers to harvest the victims' credentials. 
In this type of scam, the phishers also used domain names, which contained Meta related words (Meta, FB, Facebook, Instagram, etc.), such as domain \texttt{instagramsupport.net}, as well as words related to the concept of the phishing campaign. For example, they used the word 'support' to trick their potential victims/campaign recipients into clicking a site that resembled Meta's support page, which would help them secure their accounts.

\begin{figure}[th]
    \centering
    \includegraphics[width=.45\textwidth]{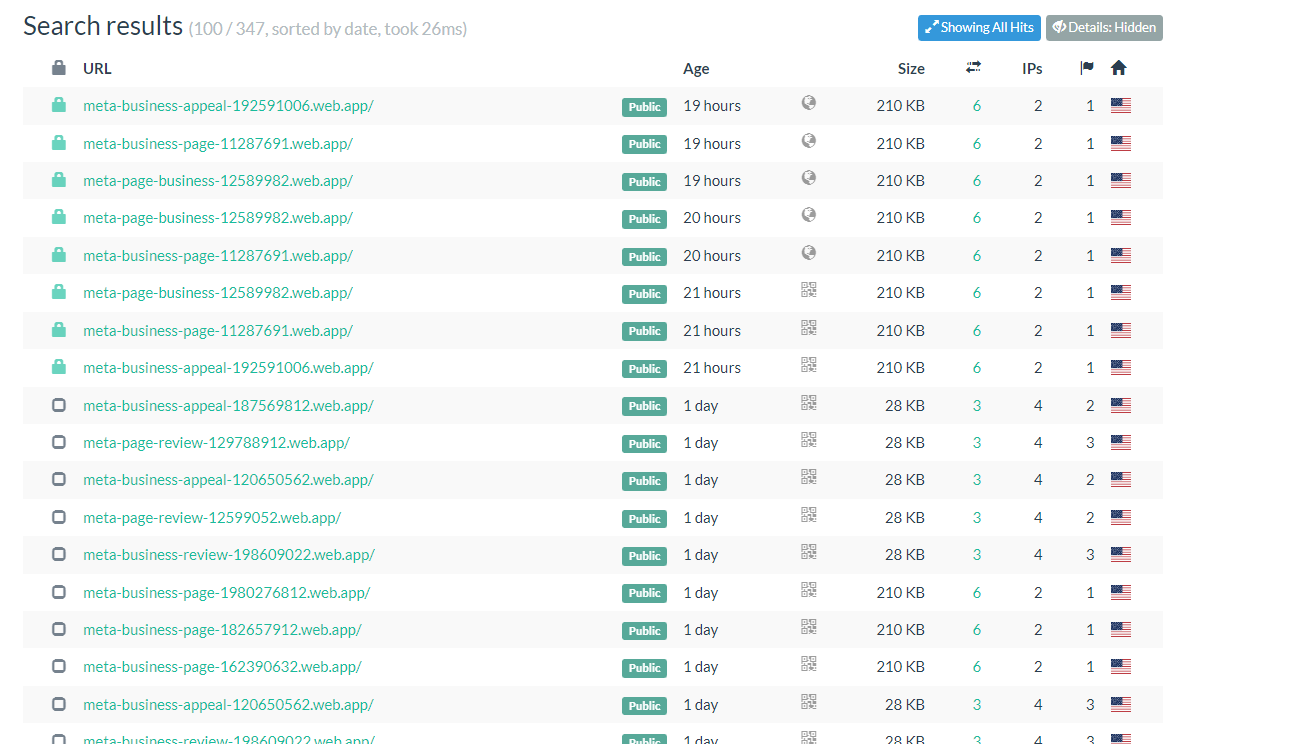}
    \includegraphics[width=.45\textwidth]{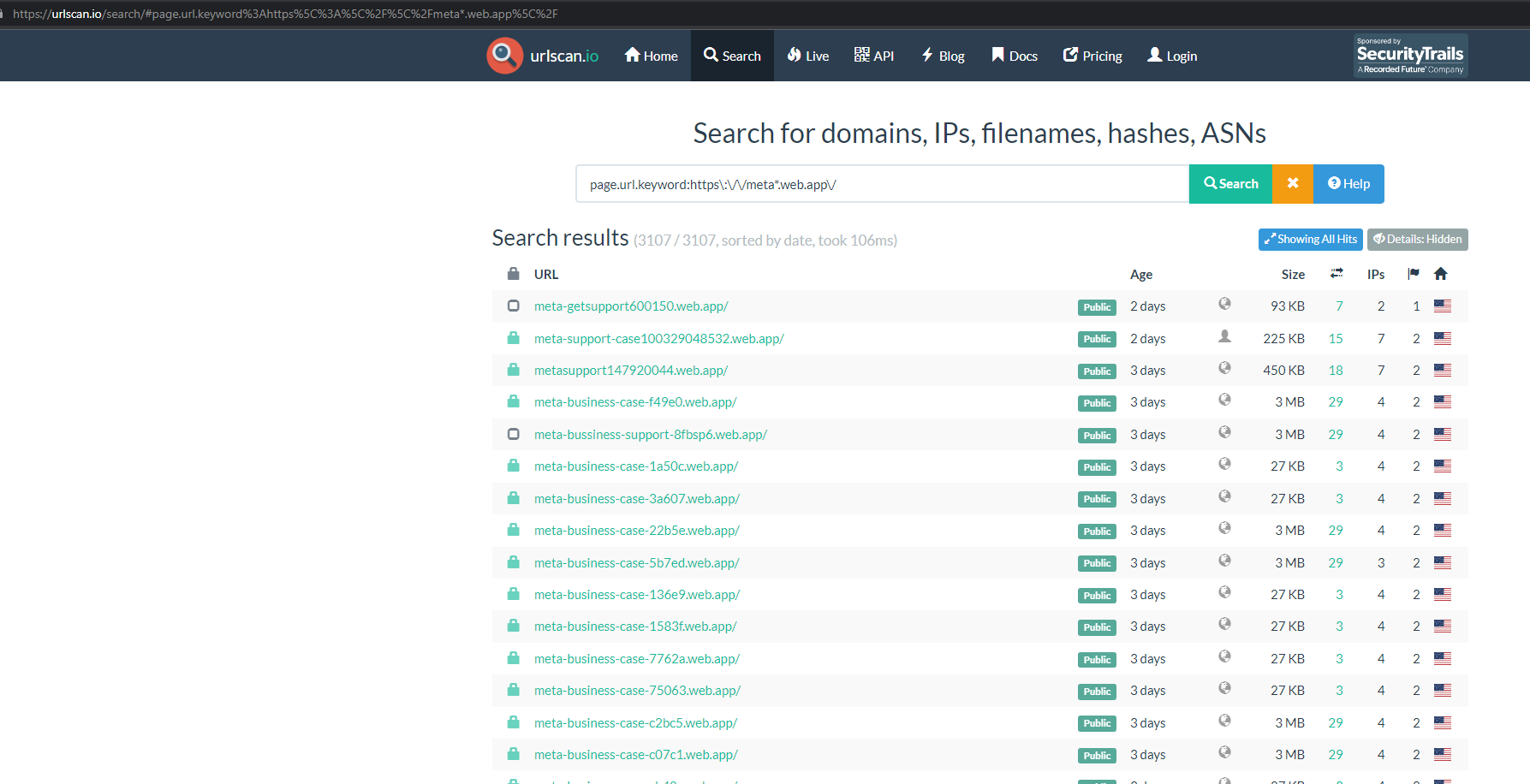}
    \caption{Searches on URLScan.io for meta*.web.app websites.}
    \label{fig:urlscan}
\end{figure}

In another observed phishing campaign concept, the phishers informed potential victims that they had infringed a trademark. Thus, the potential victims should visit the relevant supposed Meta page to object to the supposed relevant decision or to state their opinion on the alleged infringement. They also needed to do so before Meta reached a respective decision and possibly closed their Meta account. In this type of scam, the phishers seem, the latest from September 2022, to favour popular and reputable web hosting services, such as \texttt{Google Firebase} and \texttt{.web.app} TLD, to host their phishing sites. Phishers use domain name patterns such as \texttt{meta-business-appeal*.web.app}, \texttt{meta-page-review*.web.app}, \texttt{meta-page-appeal.web.app}, etc. Note that \texttt{*} can be replaced with numbers or extra letters. Data from OSINT search engines such as \texttt{Securitytrails URLScan.io}\footnote{\url{https://urlscan.io/}} show that on 25/9/2022, 347 searches were performed for sites that matched the pattern \texttt{https://meta*.web.app/} and on 17/6/2023 the number had risen to 3107. A sample of \texttt{Securitytrails URLScan.io} performed URL searches can be found in Figure \ref{fig:urlscan}.

Various trends regarding the location where victims' data were sent were identified. Phishers beyond merely storing the data on the backend of the phishing page or sending the data to them using traditional email mechanisms, abused Telegram bot API and Firebase Storage.

\begin{figure}[th]
    \centering
    \includegraphics[width=.48\textwidth]{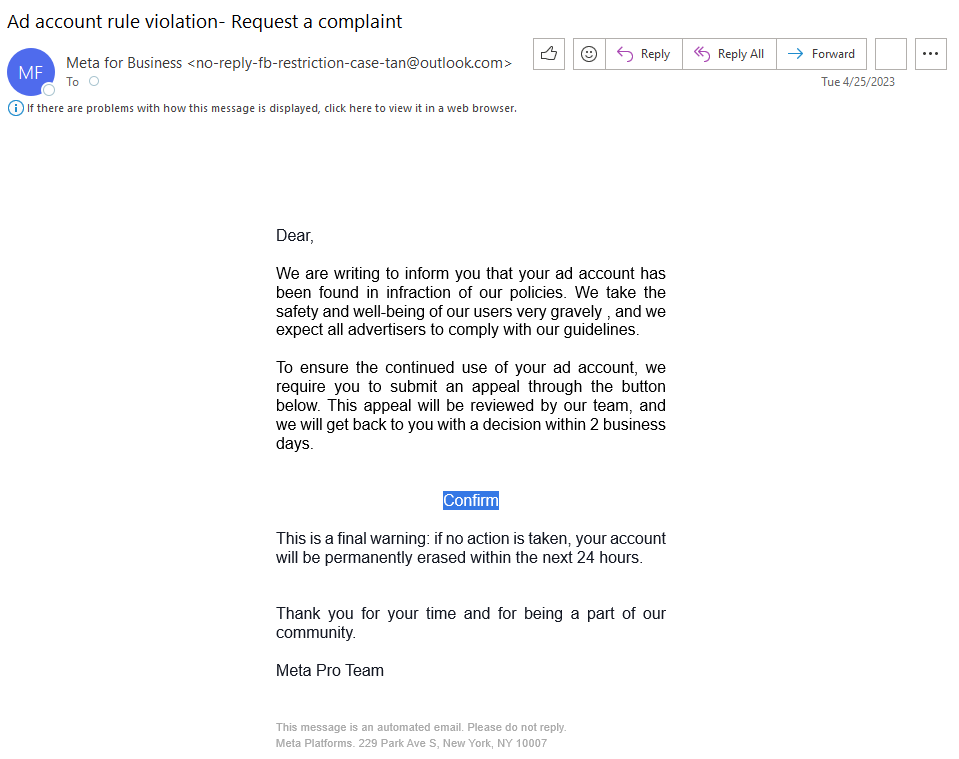}
    \includegraphics[width=.48\textwidth]{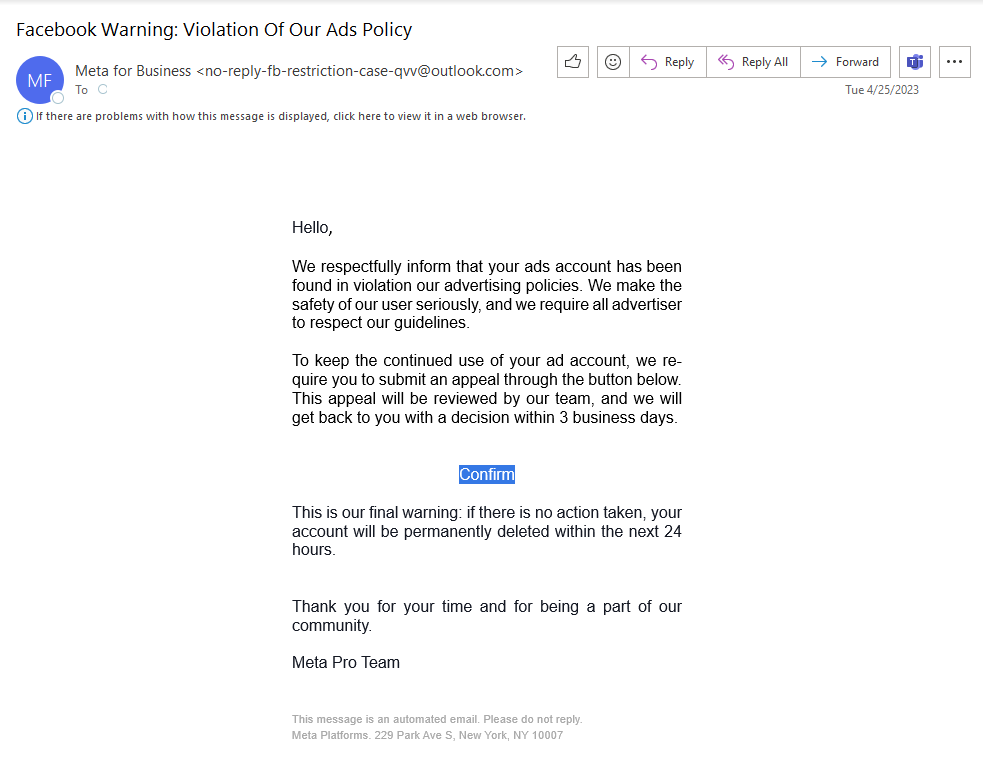}
    \caption{Campaign's phishing email.}
    \label{fig:emails1}
\end{figure}

\begin{figure}[th]
    \centering
    \includegraphics[width=.45\textwidth]{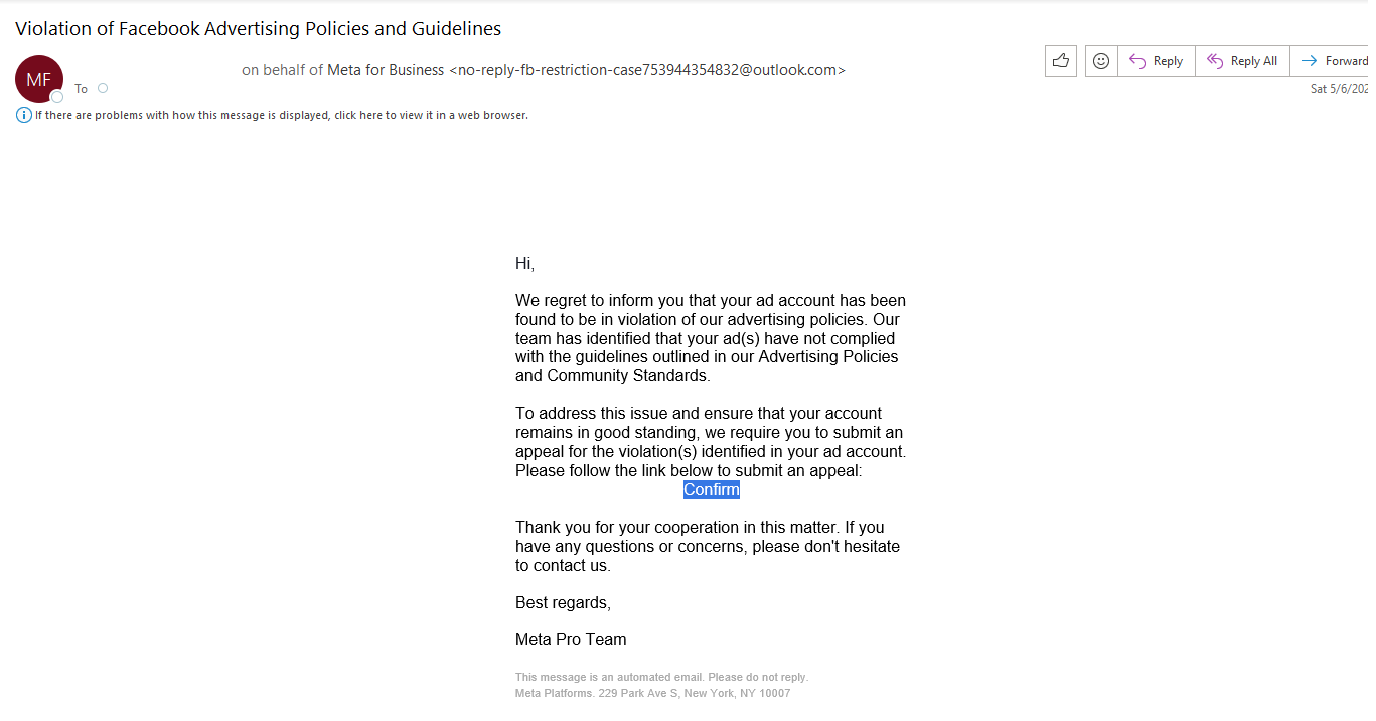}
    \includegraphics[width=.45\textwidth]{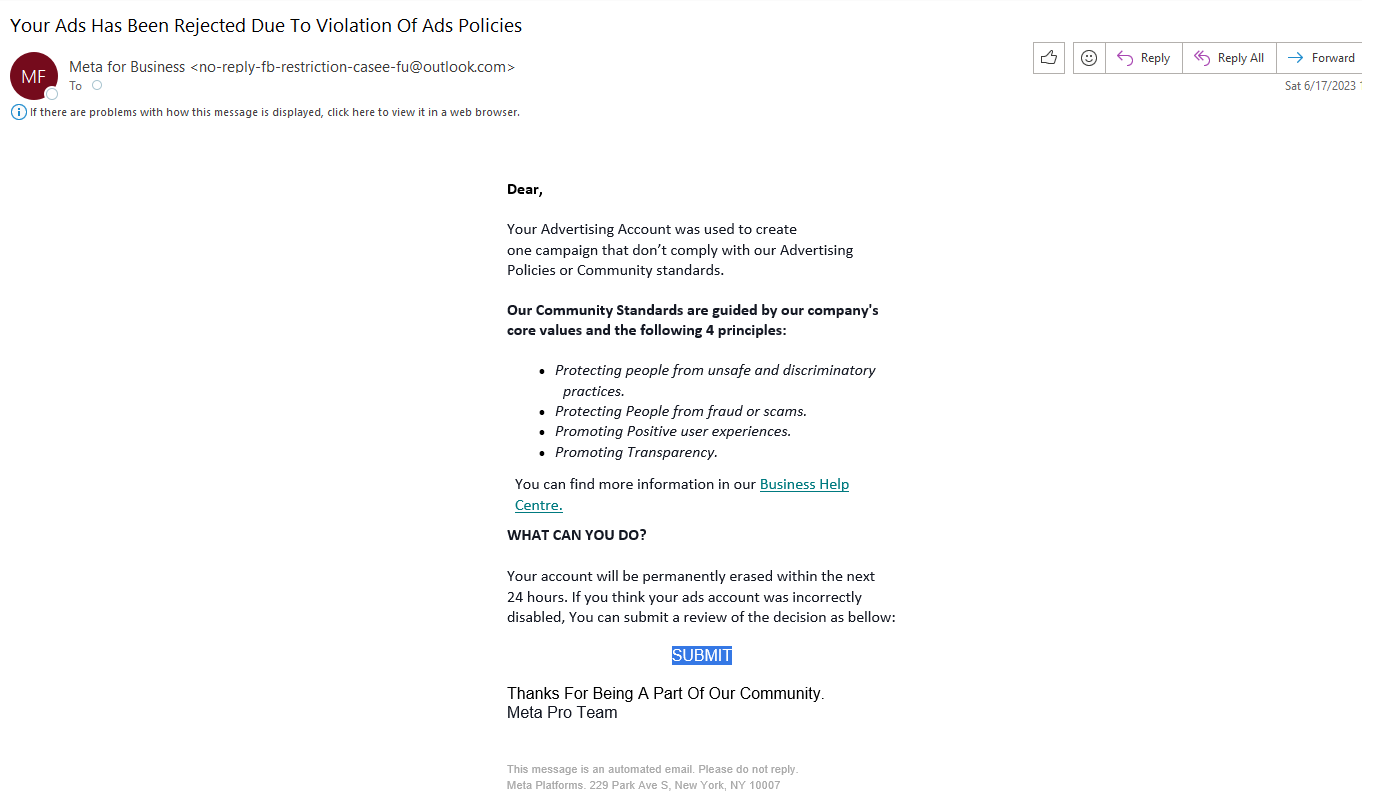}
    \caption{Campaign's phishing email.}
    \label{fig:emails2}
\end{figure}

Moreover, we observed a large ongoing set of campaigns, which targeted Meta in the last seven months. The phishers utilise mostly the \say{Advertising Policies} violation phishing campaign concept and employ Google Firebase \texttt{.web.app} TLD, along with an email-sending service to send, process, and store the victim's data. Some examples of utilised phishing sites are of the form \texttt{business-confirm-appeal-*.web.app}, \texttt{fb-restriction-cas*.web.app},  \texttt{facebook-help*.web.app}, \texttt{facebook-help*.web.app}, \texttt{business-confirm-request*.web.app}, \texttt{business-restriction-cas*.web.app}, \texttt{ad-account*.web.app}, \texttt{business-appeal-form*.web.app}, \texttt{due-to-policy*.web.app}, etc. At the time of writing, we know more than 6500 sites that can be attributed to this set of campaigns and the malicious actors behind them. This number comes as a result of analysing multiple emails that were sent as part of this set of campaigns and included respective phishing sites, as well as by using OSINT sources and exploiting technical and operational deficiencies. Figures \ref{fig:emails1} and \ref{fig:emails2} show two samples of these phishing emails, which were received by unwitting potential victims. The analysis allowed us to identify commonalities and recognise a common modus operandi, which points to a group of malicious actors being associated with the phishing websites and the associated campaigns. More about this set of campaigns and the associated with it collected data will be discussed in the following sections. 

It should be noted that the annual study by Interisle Consulting Group\footnote{https://interisle.net/PhishingLandscape2023.pdf} on the current phishing landscape reports that Meta's Facebook is the second most targeted brand, with the third one is the United States Postal Service. At the same time, the study reports Google (firebaseapp.com, web.app TLD) as the second-ranking hosting provider used by phishers in the last year. The report's findings are fully aligned with the phishing trends that we have identified in our research. 

\subsection{Sentiment analysis}
As Internet-based applications like social media platforms and blogs heavily expanded and became part of everyone's way everyday activity, daily activity-related comments and reviews are being heavily produced on the Internet. Sentiment analysis, also known as opinion mining, is the process of compiling and examining people's comments and reviews, which contain opinions, ideas, and impressions about various entities, such as goods, topics, organizations, services, as well as their attributes \cite{liu2012sentiment}. Sentiment analysis uses natural language processing and text mining techniques to identify and extract insightful information from a text passage. It utilises this extracted information to classify the text passage into positive, negative, or, occasionally, neutral categories \cite{sanchez2019social}. As the Internet can be used to gather a trove of information which resides in various sources, such as social media, news sources, e-commerce websites, forums, weblogs, and other websites, sentiment analysis can be applied to many different fields, effectively attracting not only researchers but corporations, governments, and other organisations, always depending on the context of the collected information and the goal of the intended analysis \cite{sanchez2019social}. In the next paragraphs, the main process for sentiment analysis will be adequately described, as well as the models used in performed research for sentiment analysis.

After collecting the data through web scraping from the internet, a researcher can focus on the feature selection and extraction stage, which significantly impacts the sentiment analysis model's performance. Some helpful context features are slang words, emojis, and punctuation marks, while stylometric features, such as word length, sentence length, and word/character N-grams \cite{ahuja2019impact, agarwal2011sentiment} could also be considered. Feature extraction tries to extract insightful characteristics that capture the text's most crucial elements. Two of the simplest approaches for extracting text features are methods (a) bag of words (BoW)  and (b) term frequency-inverse document frequency (TF-IDF)  \cite{sparck1972statistical}. The BoW method counts the number of times that each word appears in a given text, whereas TF-IDF provides each word with a weight. We can distinguish between significant terms and frequently used words by utilizing TF-IDF. Word counting is a component of both approaches. Word embeddings constitute another technique for feature extraction, represented by a real word vector that encodes similar-meaning words with similar embeddings. Models which utilise word embeddings are Word2Vec \cite{mikolov2013efficient}, paragraph vectors \cite{le2014distributed} also known as Doc2Vec, GloVe \cite{pennington2014glove}, FastText \cite{bojanowski2017enriching} and ELMo \cite{peters2018semi}. Additionally, there are word embeddings specifically designed for sentiment analysis \cite{tang2015sentiment}. Each of these embedding models uses a vector with a specific dimension to attempt to represent the meaning of a word. The classification models, which are covered later on, are built upon the input representations presented in this section. 

There are mainly three approaches for sentiment analysis; (a) lexicon-based, (b) machine learning, and (c) hybrid approaches. Lexicons are collections of tokens, each with a predetermined score indicating the text's neutral, positive, or negative sentiment \cite{kiritchenko2014sentiment}.  An analysed document is first separated into single-word tokens, after which the polarity of each token is determined and then summed. Polarity is the extent to which the text expresses a positive or negative sentiment. Tokens are assigned a score either based on their polarity, such as +1, 0, or -1 for positive, neutral, or negative, or based on their degree of polarity, with values ranging from +1 to -1, where +1 denotes a highly positive state and -1 a strongly negative state \cite{wankhade2022survey}. The primary disadvantage of lexicon-based approaches is that they are heavily domain-focused, and terms from one domain cannot be used in another \cite{moreo2012lexicon}. Thus, careful domain consideration should be made when assigning polarity to words. Machine learning approaches, on the other hand, use syntactic and/or linguistic features to understand patterns and solve sentiment analysis tasks. In machine learning models, sentiment analysis relates to performing a standard text classification task using given labels. Traditional machine learning models, such as Support Vector Machines \cite{boser1992training} and Decision Trees \cite{quinlan1986induction}, have been used by researchers to perform sentiment analysis tasks. Neural networks mainly outperform traditional machine learning models by capturing more complex relationships between the labels and the data. These models use word embeddings in the text sequence as input and provide a fixed-length vectorial representation of the text's meaning. Recurrent Neural Networks (RNNs) \cite{liu2016recurrent} and their variants, such as Long Short-Term Memory (LSTM) \cite{bandara2020forecasting}, GRU \cite{cheng2020text} and Bi-LSTM \cite{abid2019sentiment}, became significantly known due to their high performance and use in various NLP tasks including but not limited to sentiment analysis. Convolutional neural networks have also been used in sentiment analysis and tested in combination with other models, such as RNNs \cite{basiri2021abcdm}. Recently, attention architecture has been introduced and achieved exceptional results \cite{vaswani2017attention, daniluk2017frustratingly} in numerous applications. BERT \cite{devlin2018bert}, RoBERTa \cite{liu2019roberta}, and DistilBERT \cite{sanh2019distilbert} constitute models that are based on attention architecture. Finally, hybrid approaches combine lexicon-based and machine learning techniques. Such approaches are useful when data is limited, and machine learning models may not be properly fine-tuned \cite{severyn2015twitter}. Hybrid models have been created based on SentiWordNet dictionary \cite{baccianella2010sentiwordnet} and SVM method \cite{varghese2013aspect,krishna2017feature,devi2016feature}. Other models \cite{vo2015target,bao2019attention}, which are respectively based on the Word2Vec word embedding and attention-LSTM models, incorporated sentiment lexicons at a later stage to improve their performance. 

\subsection{Emotion analysis}
Emotions are generally represented in discrete and dimensional forms, whereas discrete representations categorise emotions into finite groups. Ekman divides emotions into six fundamental categories, namely joy, anger, fear, sadness, disgust, and surprise \cite{ekman1992there}. Moreover, he posited that these emotions are independent of each other, are basic, and can produce, when combined, complex emotions. Similar to Ekman, Plutchik suggests few basic emotions with opposite pairings, which could be combined to produce complex emotions \cite{plutchik1980general}. Eight more important emotions, including joy/sadness, trust/disgust, anger/fear, and surprise/anticipation, were added by Plutchik to Ekman's taxonomy. According to Plutchik, each emotion's intensity fluctuates depending on how the respective person interprets a situation. The OCC paradigm \cite{ortony2022cognitive}, proposed by Orthony, Clore, and Collins, states that emotions form based on people's perceptions of events and that their strength can fluctuate. OCC paradigm rejects the concept of Ekman and Plutchik, which, as stated above, divides emotions into \say{fundamental emotions}. Recently, a new emotion taxonomy model, which is called GoEmotions, has been proposed. This model departs from the Ekman taxonomy and uses a cutting-edge taxonomy of 27 categories for emotions as well as one neutral category. With the help of this expanded taxonomy, emotion analysis models work better and cover a wider spectrum of emotions \cite{demszky2020goemotions}. Comparatively, dimensional models classify emotions into one or more dimensions. This arrangement illustrates the correlation between emotional intensity and frequency. Indicative examples of two- and three-dimensional frameworks are presented in \cite{russell1980circumplex, russell1977evidence}, as well as in \cite{cambria2012hourglass}, where the Hourglass of Emotions model framework was first introduced, while the latter framework was improved in \cite{susanto2020hourglass}. 

Emotion analysis and sentiment analysis are two closely related disciplines. While sentiment analysis frequently assigns polarity labels or scores to texts, a variety of emotions are considered in emotion analysis \cite{cambria2017affective}. Movie reviews have been analysed and mapped with emotion scores based on the Hourglass of Emotions dimensions \cite{topal2016movie, cambria2012hourglass}. In the Aspect-Based Sentiment Analysis field, emotion analysis has been used in several studies, in tasks such as analysis of restaurant reviews \cite{suciati2020aspect}. Additionally, it is possible to conduct a mixed sentiment and emotion analysis. For instance, Weichselbraun et al. \cite{weichselbraun2017aspect} presented a methodology that enables the extraction of affective knowledge at the aspect level, considering both sentiment polarities and emotion categories. Emotion analysis becomes more difficult when code-switched text is analysed, as the emotion in such a text is difficult to predict. By combining both monolingual and bilingual information, a bilingual attention network model was created, which attempted to capture the emotions present in code-switching texts. An attention mechanism was created that was used to recognise important words from both monolingual and bilingual settings, while an LSTM model was used to build a thorough representation of each analysed post at the document level \cite{wang2016bilingual}. In another work \cite{zhou2018emotional}, an emotional chatbot is described, which uses GRU to simulate how emotions affect the creation of widespread discussions. A similar approach has also been used in other sentiment analysis works, such as \cite{tang2015document, zhang2016gated, zhang2016tweet, abdul2017emonet}. In another work \cite{felbo2017using}, researchers used millions of emoji occurrences derived from social media to build pre-trained neural models to enhance the representation of emotional circumstances. A question-answering approach for emotion cause extraction was used to perform emotion analysis, with researchers \cite{gui2017question} using a deep memory network to extract the reasons behind emotions that are stated in a given text. Numerous disciplines, including dialogue utterances \cite{huang2019emotionx,huang2019ana}, cyber abuse \cite{malte2019multilingual}, fake news and propaganda \cite{vlad2019sentence}, and personality traits \cite{kazameini2020personality,mehta2020bottom}, have utilised transformer architecture models.

\section{The phishing campaigns in focus}
As discussed, this work focuses on specific phishing campaigns for Meta users. The phishers behind these campaigns sent emails through a trusted mail transfer agent (MTA), namely \texttt{Salesforce}. In the body of the email, the phishers reveal their target, which is the recipient's Meta advertising account, which is referred to as \say{ad account}, \say{advertising account}, or \say{ads account}. 

The use of a trustful MTA is a method that is often utilised by phishers to maximise their chances of passing email filters, as it minimises the chances of their emails being flagged as spam or malicious. Practically, this provides their emails with the proper SPF, DKIM, and DMARC headers, providing the infrastructure to send emails to a huge number of recipients, even systematically, through the use of APIs, but also monitoring who, when, and from where opens the emails. 

The messages entail a level of seriousness and play with the victims' emotions as they inform them that the alleged sending entity (Meta) considers its users' safety to be of great importance and needs them to respect Meta's guidelines. The victims have supposedly not respected these guidelines and will soon be \say{punished}; e.g., their Meta account will be \say{permanently erased within the next 24 hours}. The only way to avoid this is for the victims to appeal to the supposed Meta's decision by visiting the sites included in the emails. In this way, victims are pressed even more as the time constraint is added, and they need to act before they lose the ability to advertise themselves and their businesses, as well as the corresponding Meta account. Note that since these accounts are professional, the recipients of these campaigns have invested a lot of time and effort to build their network and reputation. 

\begin{figure}[th]
    \centering
    \includegraphics[width=.8\textwidth]{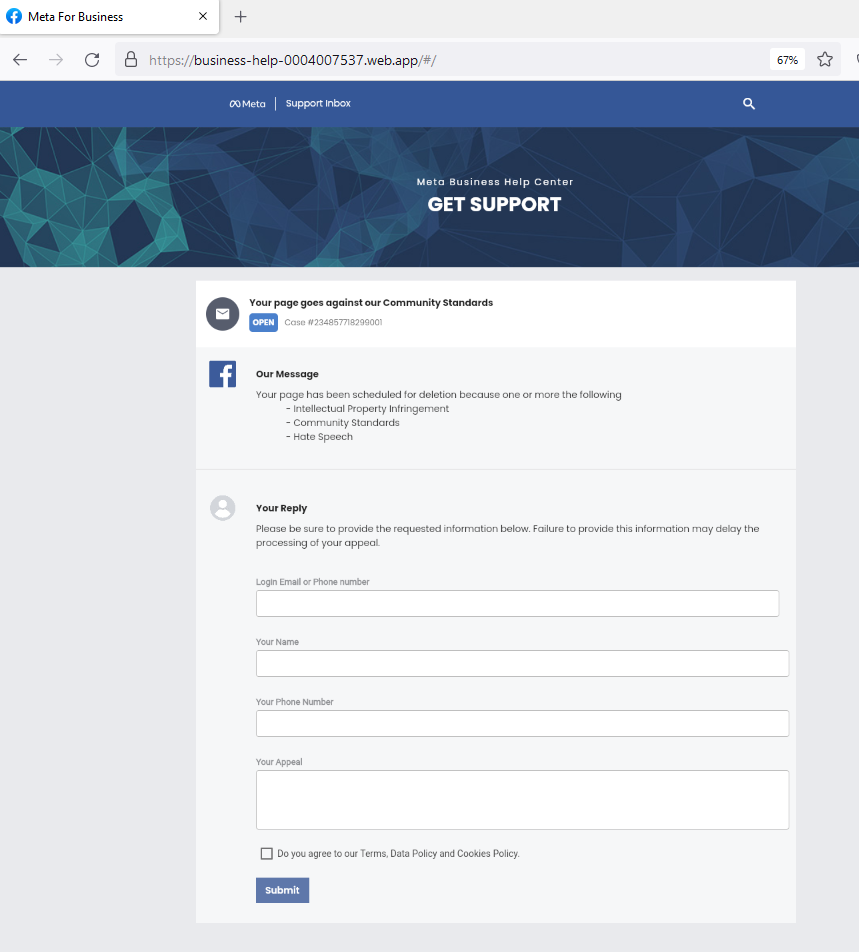}
    \caption{Meta phishing page.}
    \label{fig:phishing_site1}
\end{figure}

The phishing sites, in the latest form used, as illustrated in Figure \ref{fig:phishing_site1}, request from them that they provide their login email, name, phone number, the reason for appeal, password, as well as a two-factor authentication code. The latter two (password and TFA) are used for secure authentication/identification of a Meta user and are asked two and three times, respectively. Asking for a password and / or a TFA more than once is a common practice used by phishers, as victims tend, for example, to provide more than one real password, which they actually use (in the mimicked platform they see on the phishing site or elsewhere), when they are told that they entered the wrong password.

\begin{figure}[th]
    \centering
    \includegraphics[width=\textwidth]{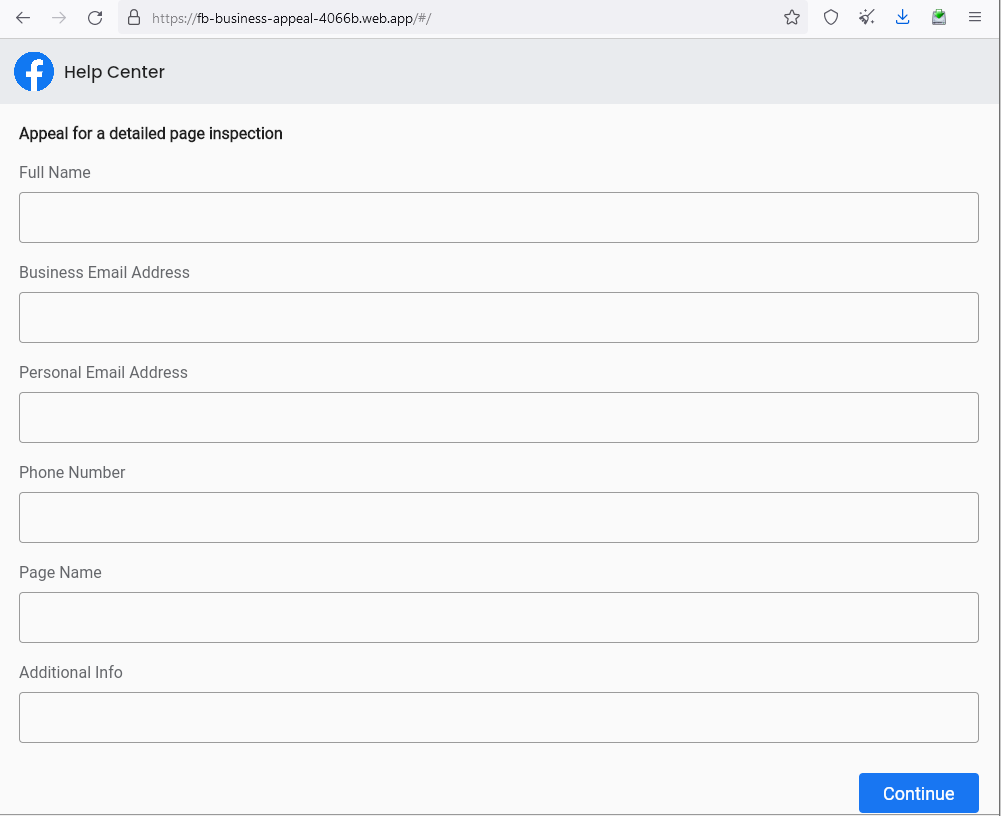}
    \caption{Meta phishing page.}
    \label{fig:phishing_site2}
\end{figure}

In their first observed form, as illustrated in Figure \ref{fig:phishing_site2}, the victims were also asked to enter their personal or business email, as well as their Facebook page name.

Based on the above, the campaigns abused three legitimate service providers, providing the necessary quality guarantees for each part of the campaigns. The delivery used Salesforce, which guaranteed email filter bypass, and the hosting used Google's infrastructure which allowed for a robust and highly available server with a trusted certificate. Finally, the bait was the abuse of Meta policies by cloning its logos and user interface. Finally, it should be noted that the campaigns focused on professionals and not individual users of Meta, implying that the phishers already had a curated list of emails of their potential victims. Therefore, we can safely assume that the goal of the campaigns is to harvest user credentials to lock them out of their accounts and then require some monetary exchange to return them while simultaneously abusing their accounts by initiating ad campaigns that are charged to their victims, and reach out to their network.

\section{Victims' statistics}
\label{sec:vic_stats}
Due to operational issues, the phishers allowed remote read access to the backend of their infrastructure, allowing us to collect the information of their victims. As a result, we responsibly disclosed the event and relevant information to the three companies whose services were abused to take down the phishing pages, prevent further people from being exploited, and notify victims.

\begin{figure}[th]
    \centering
    \includegraphics[width=.8\textwidth]{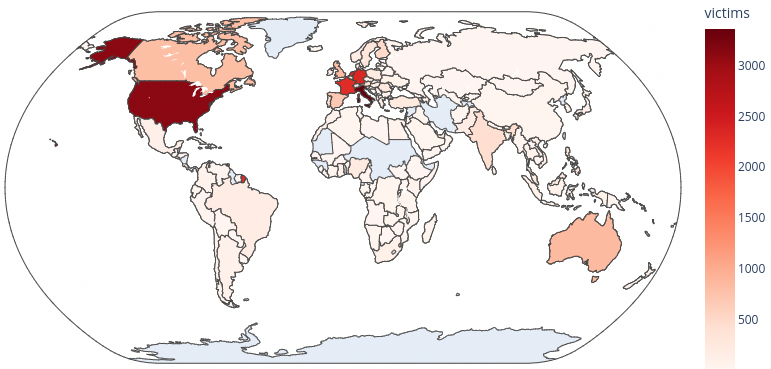}
    \caption{IPs per country.}
    \label{fig:vic_countries}
\end{figure}

From the collected data related to the aforementioned campaigns, we extracted much information about the victims. More precisely, there are 25205 unique victim emails. Based on their IP addresses, which are recorded by the phishers through their backend, we observed 24923 unique IP addresses. Using the IP Geolocation API\footnote{\url{https://ip-api.com/}}, we queried each IP to find the corresponding countries. The results indicate that the campaigns span all over the world, as the dataset contains IP addresses from 175 different countries, as illustrated in Figure \ref{fig:vic_countries}. Most IPs are from Italy (3362), the USA (3362), the Netherlands (2350), Germany (2346), and France (2282), so the victims are expected to reside in these countries. Interestingly, most users seemed to be using their mobile phones when filling these forms, using Chrome as a browser; see Table \ref{tbl:victim_stats}. Moreover, based on Table \ref{tbl:victim_stats}, we can observe a higher susceptibility of Apple users, given the significantly smaller share of Apple devices compared to Android phones and Windows hosts. The latter can be attributed to the fact that victims are handling the social media accounts of their companies, hence belong to digital marketing and creative sectors, which historically favour Apple products.

\begin{table}[th]
    \centering
    \footnotesize
    \begin{subtable}[t]{0.32\textwidth}
    \begin{tabular}{ll}
\toprule \textbf{Browser} & \textbf{Victims} \\ \midrule
    Chrome& 13618\\
 Safari& 8423\\
 Edge& 1859\\
 Firefox& 1415\\
 Facebook& 1210\\
 Opera& 176\\
 Apple WebKit& 24\\
 YaBrowser& 8\\
 UCBrowser& 2\\
 unknown& 1\\
 curl& 1\\
 Konqueror& 1 \\\bottomrule
      \end{tabular}
      \caption{Browser statistics}
      \end{subtable}
      \begin{subtable}[t]{0.32\textwidth}
      \begin{tabular}{ll}
      \toprule \textbf{OS} & \textbf{Victims} \\ \midrule
   OS X & 12289\\
   Windows 10.0 & 8953\\
   Linux & 4772\\
     Linux 64 & 148\\
     macOS High Sierra & 88\\
     macOS Mojave & 84\\
     unknown & 84\\
     Windows 7 & 76\\
     Chrome OS & 60\\
     macOS Sierra & 43\\
     OS X El Capitan & 39\\
     Windows 8.1 & 26\\
      OS X Yosemite& 18\\
      Windows 8 & 6\\
      Curl & 1\\
      \bottomrule
    
    \end{tabular}
    \caption{Reported OS.}
      \end{subtable}
      \begin{subtable}[t]{0.32\textwidth}
      \begin{tabular}{ll}
      \toprule \textbf{Platform} & \textbf{Victims} \\ \midrule
   Windows & 9055\\
    iPhone& 8396\\
   Android & 4852 \\
   Apple Mac& 4719 \\
     Linux& 148 \\
     iPad& 73 \\
     unknown& 60 \\
     Curl& 1\\\bottomrule     
    \end{tabular}
    \caption{Platform}
      \end{subtable}
      \caption{Statistics regarding users' devices.}
      \label{tbl:victim_stats}
\end{table}

Given that users submitted their passwords, we wanted to assess their security and conformance to patterns. Since the form requested two passwords, we merged both and kept the remaining 45107 unique values. First, we tried to assess the distribution of the passwords' length. Nevertheless, we noticed lengths far exceeding what would be considered a typical password, e.g., 208 passwords contained more than 30 characters. Given that humans would not normally remember such long passwords and even password managers would not generate them, we decided to investigate those cases. A closer look into those cases revealed operational issues that the phishers had, see also Section \ref{sec:ops}. More precisely, most of these records were actually texts that had to be collected from the form and URLs and were inserted in the wrong field or curses from users who understood that this was a phishing attack. Therefore, for the rest of the analysis, we pruned all values that contained the space character, had lengths longer than 30, and were URLs. With the above criteria, we ended up with 44455 passwords.

\begin{figure}[th]
    \centering
    \includegraphics[width=\textwidth]{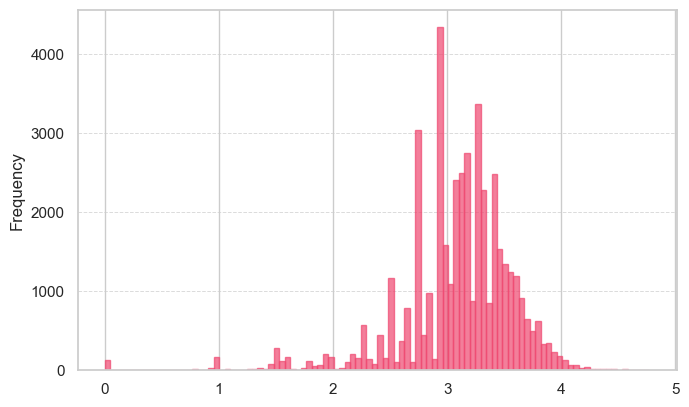}
    \caption{Histogram of the passwords' entropy.}
    \label{fig:entropy}
\end{figure}

First, we calculated the entropy of the passwords. As observed in Figure \ref{fig:entropy}, the entropy of most passwords is not optimal. Indeed, the average entropy is 3.09 bits, less than half the optimal entropy of a string that contains only printable ASCII characters, namely $\log_2 95\approx6.57$ bits per character. However, entropy is not the best way to measure password security.
\begin{figure}[th]
    \centering
    \begin{tikzpicture}
    \begin{axis}[
        xbar stacked, 
        width=\textwidth,
        height=.4\textwidth,
        bar width=15mm,xmax=45107,
        xmin=0,
        xlabel={Number of passwords},
        ytick=data,
        scaled x ticks = false,
        yticklabels={0, 1, 2, 3, 4},
        nodes near coords,
        nodes near coords align={horizontal},
        every node near coord/.append style={
            anchor=south, 
            yshift=20pt   
        },
                legend style={at={(0.5,1.2)}, anchor=north, draw=none,legend columns=-1}, 
        ]
        \addplot[col1, fill=col1] coordinates {(339,0)};
        \addplot[col3, fill=col3] coordinates {(4551,0)};
        \addplot[col4, fill=col4] coordinates {(8566,0)};
        \addplot[col2, fill=col2] coordinates {(14189,0)};
        \addplot[col5, fill=col5] coordinates {(16810,0)};
        \legend{0,1,2,3,4} 
    \end{axis}
\end{tikzpicture}
\caption{Password strength based on zxcvbn.}
  \label{fig:zxcvbn}
\end{figure}
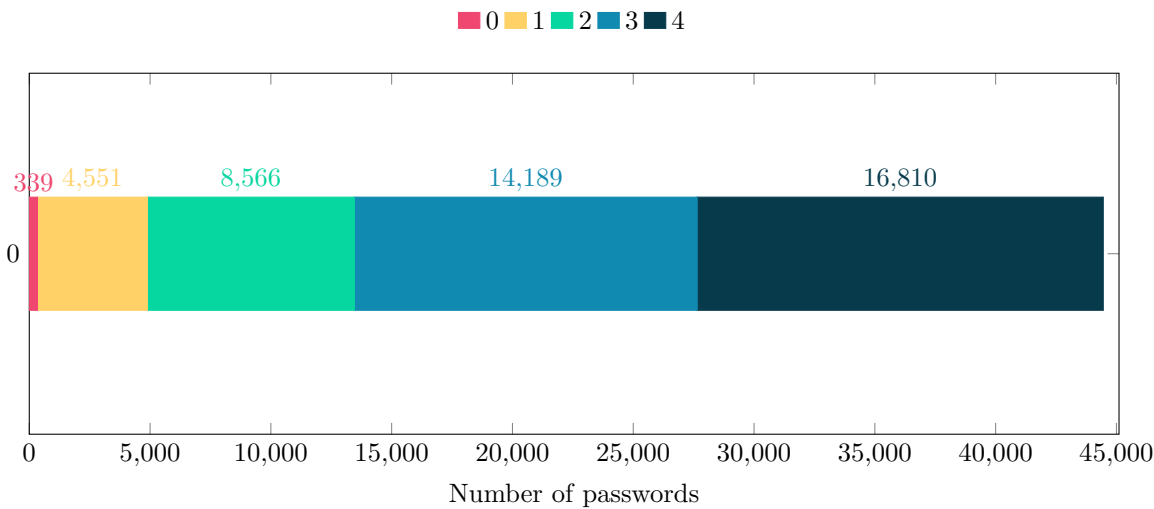

To assess the password security more accurately, we first used zxcvbn \cite{wheeler2016zxcvbn}, which measures password security on the scale of integer values from 0 to 4. As illustrated in Figure \ref{fig:zxcvbn}, while the score of most users is 3 and 4, it is clear that many users did not use very secure passwords. Notably, 30.26\% of the users reported a password with a score lower than 3. Beyond the poor choice of passwords, this also shows that Meta's password policy considers many passwords secure, even if they are not.

Next, we examined the passwords regarding previous leakage using the so-called \texttt{RockYou2021} dataset. The dataset comprises of previously leaked databases and contains 8.4 billion passwords. The dataset was initially published in RaidForums, a hacking forum seized by the U.S. Department of Justice (DOJ) in 2022; however, there are multiple mirrors online of the dataset. To our astonishment, 14676 victims' passwords were also found in the RockYou2021 dataset. Given that the leaked passwords are at least two years old, more than half of the victims (58.23\%) use leaked passwords. This staggering fact can have multiple interpretations. While users are known to be using, as previously discussed, typical passwords of low strength, the fact that these passwords have been leaked makes their choice even worse as they are susceptible to dictionary attacks. Moreover, one could also argue that these users are reusing the same password on different platforms. Thus, the phishers could use these passwords to take over other accounts of their victims, causing them further damage.

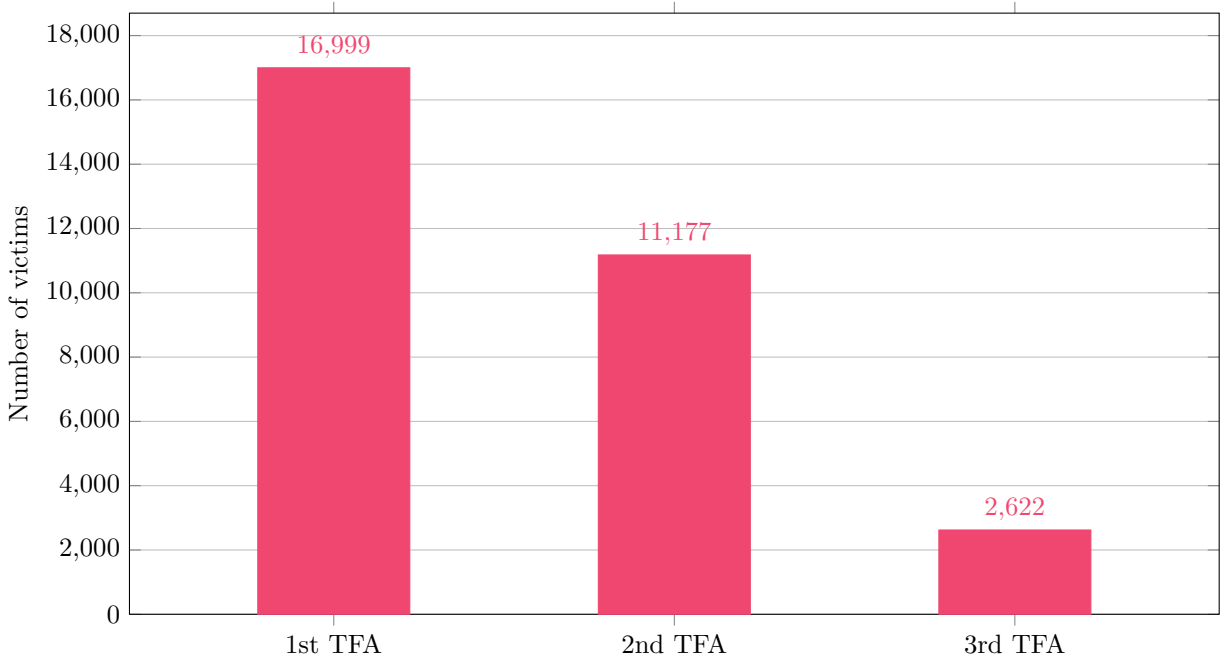
\begin{figure}[ht]
  \centering
  \begin{tikzpicture}
    \begin{axis}[
      ybar,
      width=\textwidth,
      height=.6\textwidth,
      bar width=20mm,
      ymin=0,
      ylabel={Number of victims},
      xtick=data,
      xticklabels={
        1st TFA,
        2nd TFA,
        3rd TFA,
      },
      scaled y ticks = false,
      nodes near coords,
      ymajorgrids = true,
      enlarge x limits=0.3,
      nodes near coords align={vertical},
      ]
      \addplot [col1,fill=col1]coordinates {
        (1, 16999)
        (2, 11177)
        (3, 2622)
      };
    \end{axis}
  \end{tikzpicture}
  \caption{Number of users that added 2FA. Note that users that provided the 3rd TFA have also provided the previous two, and the users that provided the second one have provided the first one too.}
  \label{fig:tfa}
\end{figure}

In Figure \ref{fig:tfa}, we report the number of users who provided the two-factor authentication numbers requested by the platform. In order to protect users from account takeovers, Meta uses two-factor authentication tokens that are sent to the user's device once the service notices abnormal activity, e.g. connection from an unknown browser, device, or country. We assert that the phishers have some automation mechanism and require the token to take over the account. Since the victims provided the credentials for their personal and professional accounts, we assume that the three tokens were requested to take over these accounts. For instance, the first token is requested to allow the login to the phisher, the second to change the password, and the third to allow changes to the professional profile. Notably, we observe that the number of users that provided these three tokens drops significantly from the first to the third token. We attribute this drop to two factors. First, the phishers, at some point, decided to remove the request for the third token. The second factor is user exhaustion and reflexes that they are doing something wrong. More precisely, the users could have stopped providing tokens as they had already provided a text to ask Meta to revise the suspension of their account, their usernames, and passwords, and additionally, they had to provide tokens that required further interaction. Moreover, we have to understand that when the request was made to Meta (from the phishers), and the token was sent to their device, the recipients would notice that Meta also had some wording on why this token was sent, e.g., password change. Therefore, while the victims fell for the bait, once they noticed contradictory messages with their tokens, their reflexes kicked in and they stopped providing further input.      

\begin{figure}[ht]
  \centering
  \begin{tikzpicture}
    \begin{axis}[
      ybar,
      width=\textwidth,
      bar width=3mm,
      ymin=0,
      ylabel={Number of victims},
      xtick=data,
      scaled y ticks = false,
      nodes near coords,
      ymajorgrids = true,
      enlarge x limits=0.025,
      nodes near coords align={vertical},
      nodes near coords style={font=\footnotesize},
      every node near coord/.append style={rotate=90, anchor=west}
      ]
      \addplot [col1,fill=col1]coordinates {
        (12, 2014)
(14, 1730)
(13, 1758)
(10, 2313)
(0, 342)
(20, 690)
(19, 883)
(18, 1017)
(11, 2232)
(15, 1618)
(16, 1488)
(22, 578)
(9, 2296)
(17, 1141)
(2, 142)
(21, 682)
(7, 957)
(23, 432)
(8, 1838)
(6, 395)
(1, 183)
(4, 129)
(5, 188)
(3, 159)
      };
    \end{axis}
  \end{tikzpicture}
  \caption{Number of victims per server time (24h).}
  \label{fig:victims_time}
\end{figure}
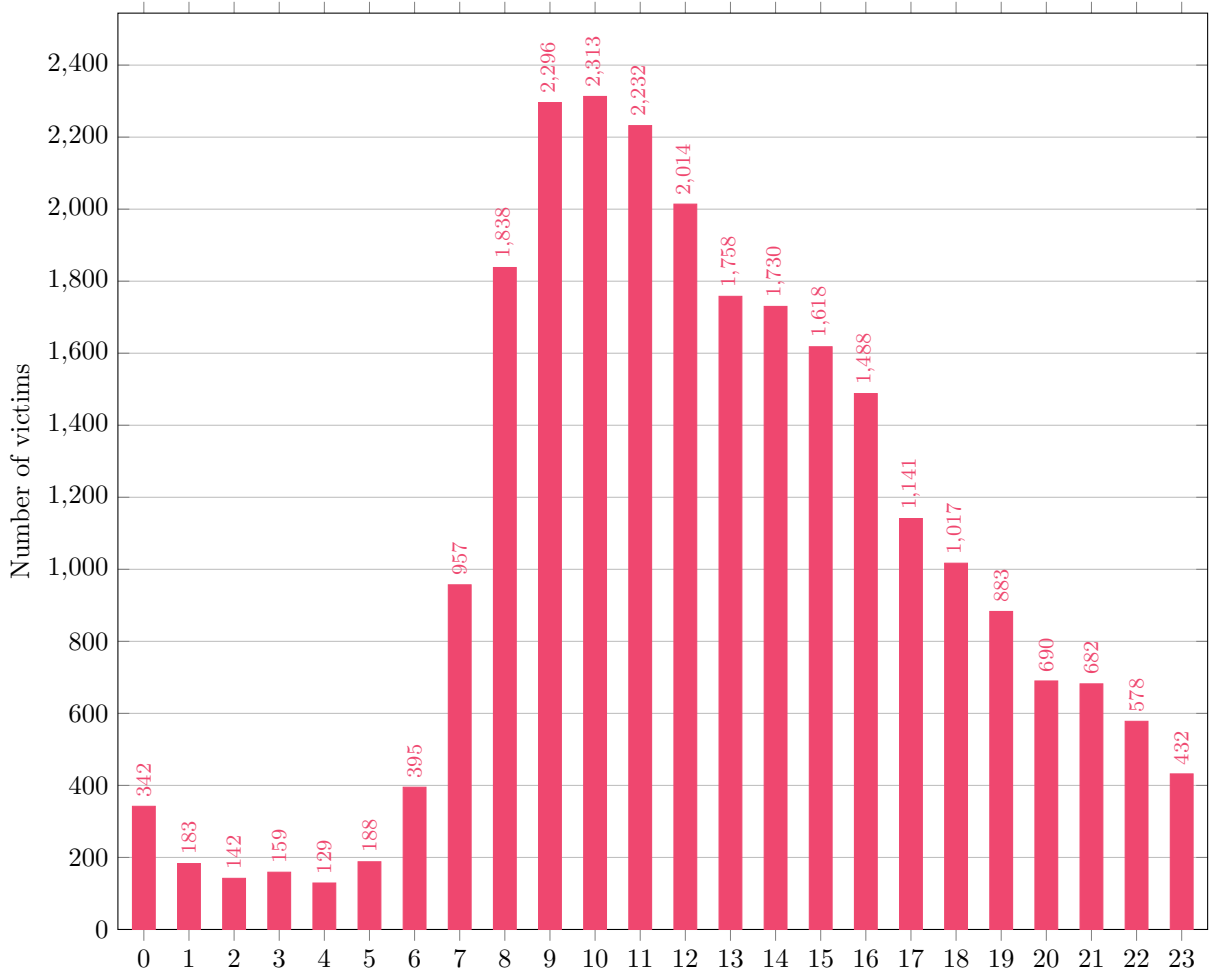

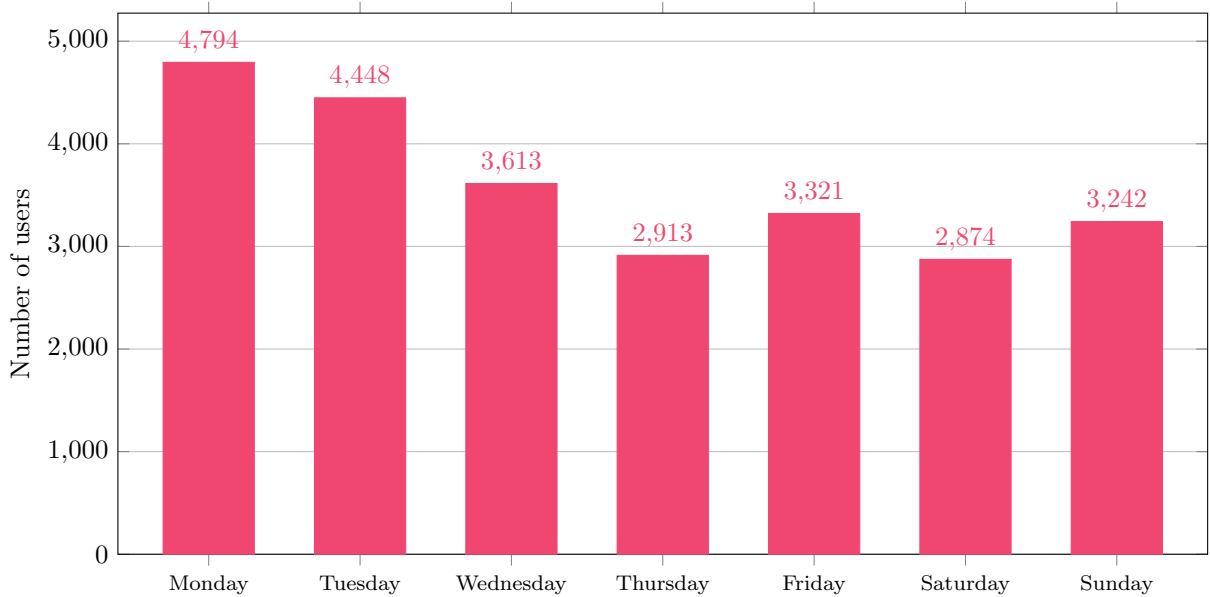
\begin{figure}[ht]
  \centering
  \begin{tikzpicture}
    \begin{axis}[
      ybar,
      width=\textwidth,
      x tick label style = {font = {\fontsize{8 pt}{10 pt}\selectfont}},
      height=.55\textwidth,
      bar width=12mm,
      ymin=0,
      ylabel={Number of users},
      xtick=data,
      xticklabels={
        Monday, Tuesday, Wednesday, Thursday, Friday, Saturday, Sunday
      },
      scaled y ticks = false,
      nodes near coords,
      ymajorgrids = true,
      nodes near coords align={vertical},
      ]
      \addplot [col1,fill=col1
      ]coordinates {
        (1, 4794)
        (2, 4448)
        (3, 3613)
        (4, 2913)
        (5, 3321)
        (6, 2874)
        (7, 3242)
      };
    \end{axis}
  \end{tikzpicture}
  \caption{Number of victim interactions per weekday.}
  \label{fig:victims_weekday}
\end{figure}

\begin{figure}[th]
    \centering
    \includegraphics[width=\textwidth]{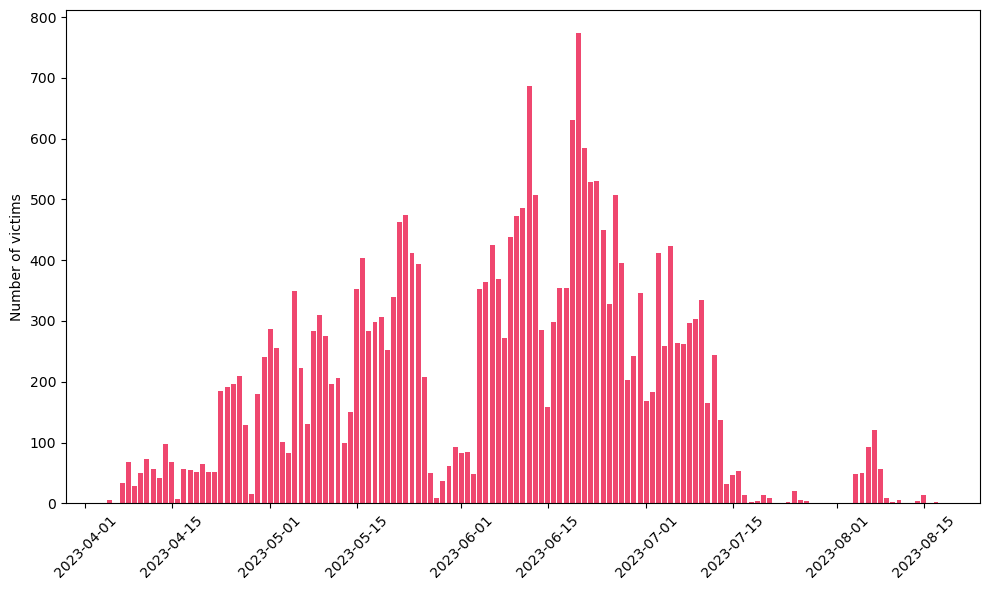}
    \caption{Evolution of the number of victims by date.}
    \label{fig:byday}
\end{figure}

\begin{figure}[th]
\centering
    \includegraphics[width=\textwidth]{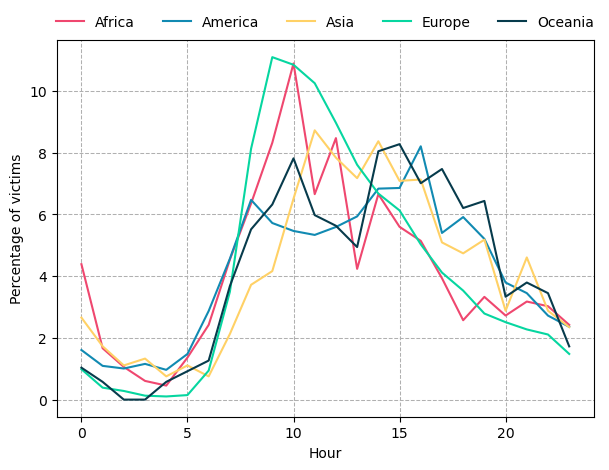}
    \caption{Distribution of victims' responses per time and continent.}
    \label{fig:bycontinent}
\end{figure}

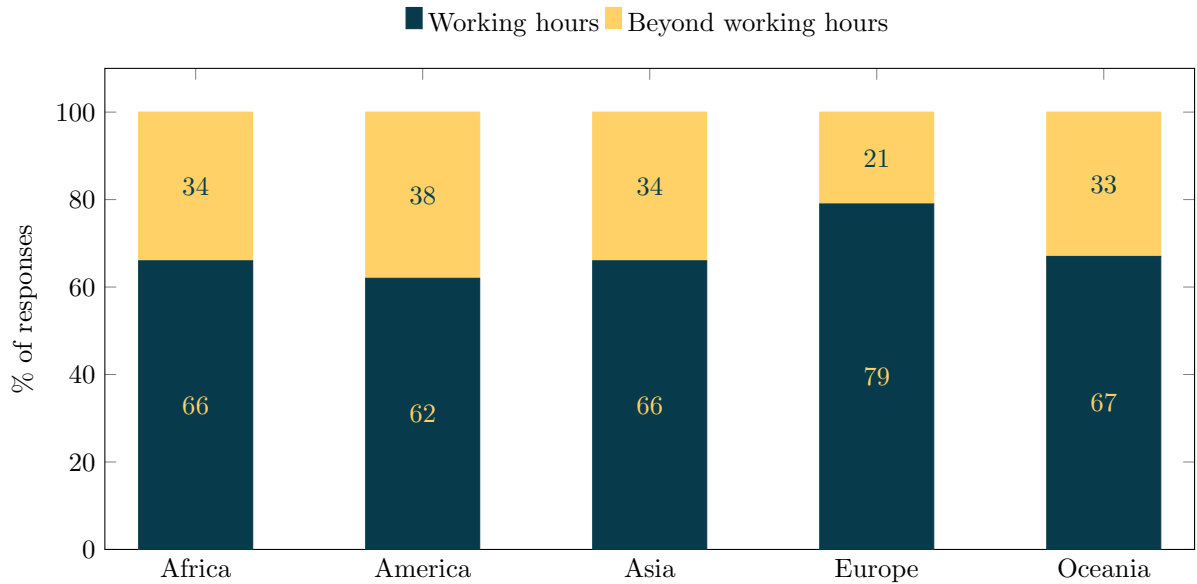
\begin{figure}[th]
        \centering
        \begin{tikzpicture}
            \begin{axis}[
                ybar stacked,
                bar width=15mm,
                ymin=0,
                width=\linewidth,
                height=0.5\textwidth,
                ylabel={\% of responses},
                nodes near coords,
                symbolic x coords={Africa, America, Asia, Europe, Oceania   },
                xtick=data,
                legend style={at={(0.5,1.14)},
                anchor=north,legend columns=-1,draw=none},
                ]
                \addplot[col5,fill=col5,text=col3] coordinates {(Africa, 66) (America, 62) (Asia, 66) (Europe, 79) (Oceania, 67)};
                \addplot[col3,fill=col3,text=col5] coordinates {(Africa, 34) (America, 38) (Asia, 34) (Europe, 21) (Oceania, 33)};
                \legend{Working hours, Beyond working hours}
            \end{axis}
        \end{tikzpicture}
        \caption{Ratio of responses within working hours and beyond them.}
        \label{fig:ratio_work_hours}
\end{figure}

Despite our continuous takedown efforts; more than 6500 domains were taken down, the number of victims reached almost 800 in a single day, as observed in Figure \ref{fig:byday}. While the campaigns are ongoing, the numbers have plunged effectively after mid July, but the phishers are gradually moving to new platforms. Although we do not have statistics on when the emails were sent, it is clear that victims responded to phishing emails on the first two days of the week, far more than on the rest of the days; see Figure \ref{fig:victims_weekday}. Finally, it is worth noting that, in terms of time, there are also patterns, as observed in Figure \ref{fig:victims_time} where we illustrate the users' responses per hour, as recorded by the phishers whose time is in UTC. Clearly, 9 UTC is the peak hour. Nevertheless, considering the IP addresses of the victims, we found a very good estimate of their timezone, which allowed us to better drill down to the response time. Figure \ref{fig:bycontinent} illustrates a more fine-grained analysis, where significant demographic differences are illustrated per continent. One can observe obvious differences in the timing of the victims' responses per continent. For instance, Europe's peak time is 9:00 AM, close to Africa (10:00 AM) and Asia (11:00 AM). However, the peak response time for victims from Oceania is 15:00, close to America (16:00 PM), while for Asia, the peak time is 12:00 PM. Given the timezone differences between the victims and the patterns we observe, we concur that these patterns do not reflect the time that the emails arrived to the victims but the differences in the mentality of the victims. A fine example is the case of an email being sent simultaneously to Oceania and America, which have several hours of difference. Furthermore, the figure illustrates the circadian rhythms of people dealing with ICT, where only the interval between 01:00 and 06:00 AM seems to be free of computer interactions, while interactions span throughout the rest of the day. Finally, as observed in Figure \ref{fig:bycontinent}, the bulk of responses are within the working hours. Nevertheless, analysing these statistics further, see Figure \ref{fig:ratio_work_hours}, it is clear that Europeans responded to the phishing campaigns by far larger extent within working hours (78.85\%). On the contrary, Americans responded more often beyond their working hours. The rest responded by approximately 66\% within the working hours.

Finally, since the backend also recorded each user response, labelling it with an ID of the corresponding phishing campaign, we investigated whether some users fell victim to more than one phishing campaign. We noticed that a daunting number of 1153 unique emails had been recorded responding to more than one phishing campaign, showing the persistence of specific users in falling victim to phishing. More precisely, 1033 unique emails had more than two recorded interactions with the phishing "platform". Since there were 120 unique emails with more than two recorded interactions, we decided to look into them better. Unfortunately, the vast majority of the 95 emails that responded to three phishing campaigns were real users; however, almost all emails with more than three responses were bogus and most likely belonged to researchers or tests of the phishers. 

This victim persistence made us investigate how often the victims returned to interact with the phishing "platform". As expected, within the timeframe of a day, we have the bulk of the users (22042). Still, there is a significant amount of users returning to interact with the "platform". More precisely, 2068 interacted twice, 371 interacted three days, 97 interacted four days, 32 interacted five days, and 11 interacted six days. The rest who interacted far more times are dummy records that belong most likely to researchers and the threat actors that tested their platform. These results signify that not only are some people prone to phishing attacks, but they are persistent enough to continue pursuing the bait even days after the attack.

\section{Text input analysis}
\label{sec:text_analysis}
In what follows, we analyse the text input that victims provided as an appeal, trying to understand how they felt and perceived the phishing email and the attack from their perspective. We enriched these responses with each user's continent and local time to determine possible correlations between sentiment, emotions, tone, demographics, and timing. 

\subsection{Dataset}
Our dataset contains texts in 41 different languages, including but not limited to English, Spanish, and German, with a total of 201138 texts from 23023 different authors, while 33544 texts are unique. It should be mentioned that all the authors, except one, have provided more than one different text. In what follows, we focus only on texts written in English. Before continuing with other dataset statistics, it is worth mentioning that there are 2346 instances that the sentiment analysis models we used could not determine. These texts contain only email and phone addresses with invalid information, which means that users realised that the message was a scam. For instance, a user provided the following contact details: \texttt{2333333333333333333333} and \texttt{scam\@scam.de}. Based on our measurements, there is no correlation between victims' demographics and text length. 

\begin{figure}[ht]
  \centering
  \begin{tikzpicture}
    \begin{axis}[
      ybar,
      width=\textwidth,
      height=.55\textwidth,
      bar width=15mm,
      ymin=0,
      ylabel={Number of text chunks},
      xtick=data,
      xticklabels={
        Africa,
        America,
        Asia,
        Europe,
        Oceania
      },
      scaled ticks=false, 
      tick label style={/pgf/number format/fixed},
      nodes near coords style={/pgf/number format/fixed},
      nodes near coords,
      ymajorgrids = true,
      nodes near coords align={vertical},
      ]
      \addplot [col1,fill=col1]coordinates {
        (1, 1070)
        (2, 7117)
        (3, 3378)
        (4, 19526)
        (5, 1442)
      };
    \end{axis}
  \end{tikzpicture}
  \caption{Distribution of text chunks per continent.}
  \label{fig:chunks_continents}
\end{figure}
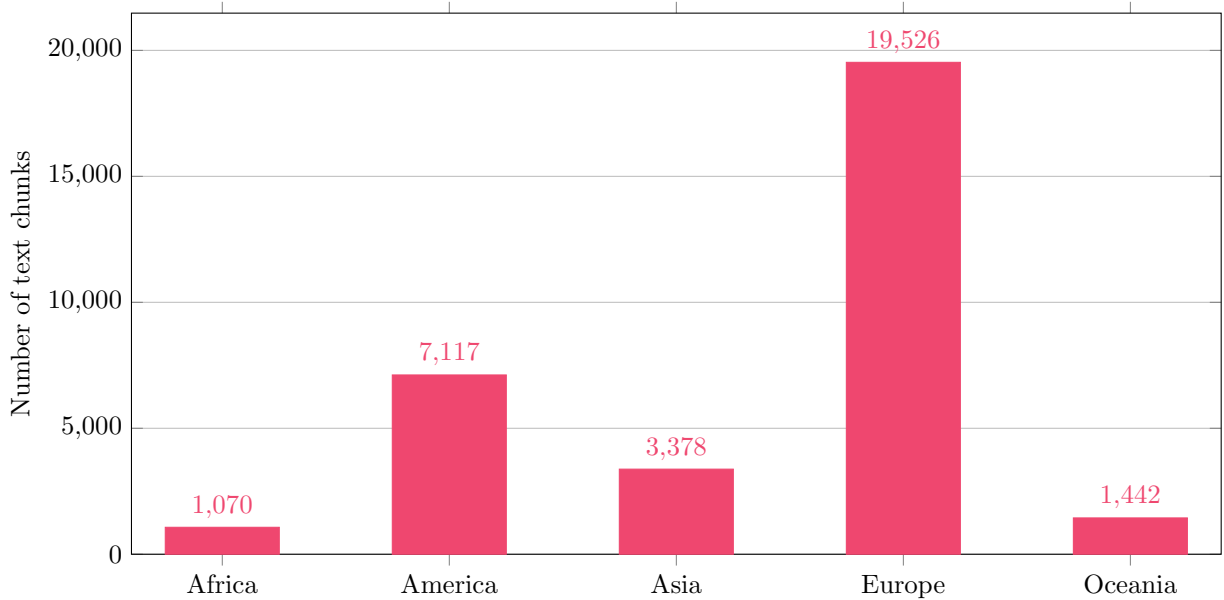

The text length is a beneficial characteristic of the data since we will use transformers, and their input size is limited. The dataset contains user-submitted objections/comments containing more than 512 characters, which is the maximum allowed input size for the used Transformers. We overcome this restriction by splitting the text into 512-character text chunks. After splitting the texts into text chunks and removing the texts containing only contact information, such as emails and phone numbers, we split them into text chunks of 512 characters. Figure \ref{fig:chunks_continents} presents the distribution of text chunks per continent. It is clear that texts from Europe far outnumber the rest of the world, and the chunks from Africa and Oceania are almost equal. The latter indicates that the Europeans were more expressive, thus eager to appeal than the others.

Table \ref{tab1:statsdataset} illustrates the statistics of the English text fragment of the dataset, which we fed into a local transformer-based model to analyse users' emotions and sentiments from their textual input. We opted for the use of local models to respect the victims' privacy. Using regular expressions, we identified that users provided additional contact information in their texts. More precisely, 364 texts contain emails, and 492 contain telephone numbers. The dataset also contains 111 emojis, where the most commonly used one is the \emph{praying hands} ((\includegraphics[height=11pt]{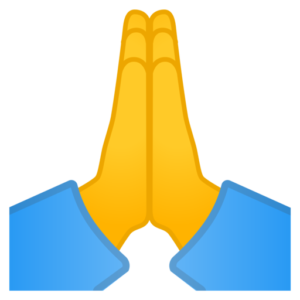})), which is often used to express gratitude and thankfulness, as well as to convey positive intentions.

\begin{table}[th!]
\centering
\begin{tabular}{lr}
\toprule
\textbf{Feature} & \textbf{Number}\\
\midrule
Authors & 18281\\
Texts & 32533\\
Words & 1411428\\
Characters  & 7937381\\ \bottomrule
\end{tabular}
\caption{Statistics of the final dataset.}
\label{tab1:statsdataset}
\end{table}

Since the users respond to a false allegation, we wanted to examine whether they would resort to swearing words. Therefore, we used two open-sourced datasets\footnote{\url{https://github.com/hpclab/DevCommunities/} and \url{https://github.com/ConsoleTVs/Profanity/blob/master/Dictionaries/Default.json}} that concern (a) a list of adult keywords, which was used in \cite{coletto2016devcom} to filter adult incoming queries landing to Tumblr blogs, and (b) a dictionary list contained within a PHP library that can block profanity words from any \say{given} string. Note that if a victim's reply contains profanity, it is not necessary that it curses the alleged sender. In our dataset, 408 texts contained profanity, using 41 unique profanity words.

\subsection{Sentiment analysis}
\begin{figure}[ht]
  \centering
  \begin{subfigure}{0.48\textwidth}
  \centering
  \includegraphics[width=\textwidth]{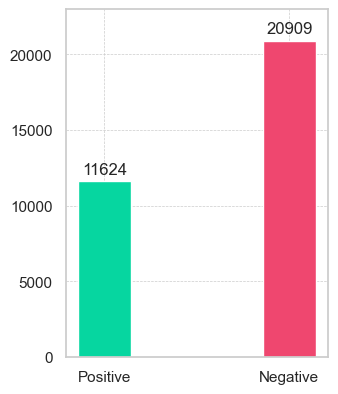}
    \caption{Distribution of text chunks per negative and positive labels.}
    \label{fig:distilbert_results}
  \end{subfigure}
  \begin{subfigure}{0.48\textwidth}
    \centering
    \includegraphics[width=\textwidth]{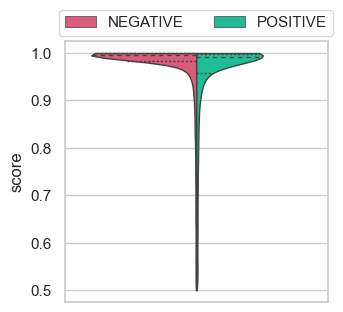}
    \caption{Distribution of scores per negative and positive labels.}
    \label{fig:distilbert_score}
  \end{subfigure}
  \caption{Results of the DistilBERT model for text chunk victims' responses.}
\end{figure}

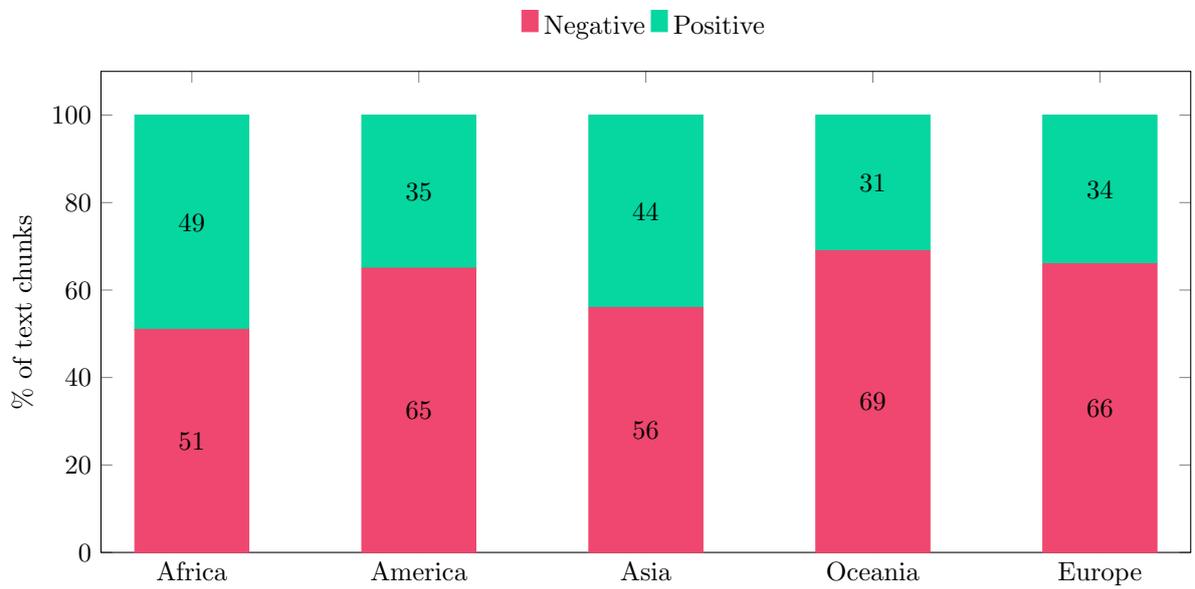
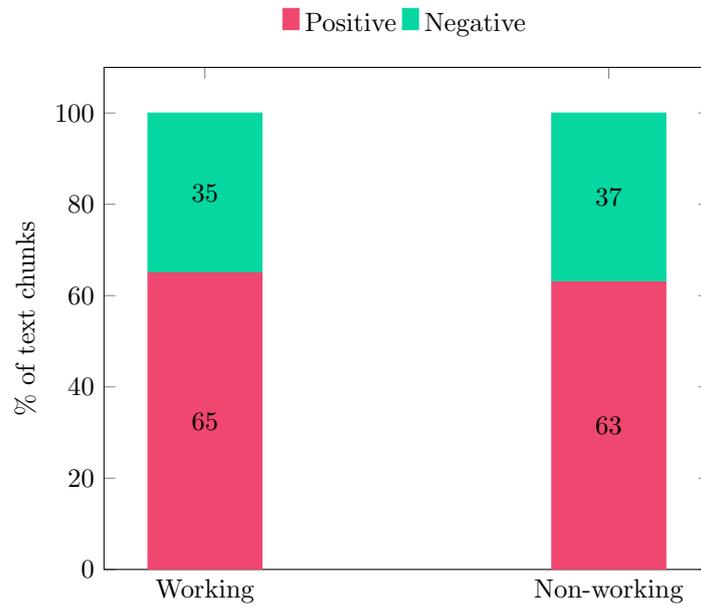
\begin{figure}[th]
    \begin{subfigure}{\textwidth}
        \centering
        \begin{tikzpicture}
            \begin{axis}[
                ybar stacked,
                bar width=15mm,
                ymin=0,
                width=\linewidth,
                height=0.5\textwidth,
                ylabel={\% of text chunks},
                nodes near coords,
                symbolic x coords={Africa, America, Asia, Oceania, Europe},
                xtick=data,
                legend style={at={(0.5,1.14)},
                anchor=north,legend columns=-1,draw=none},
                ]
                \addplot[col1,fill=col1,text=black] coordinates {(Africa, 51) (America, 65) (Asia, 56) (Oceania, 69) (Europe, 66)};
                \addplot[col4,fill=col4,text=black] coordinates {(Africa, 49) (America, 35) (Asia, 44) (Oceania, 31) (Europe, 34)};
                \legend{Negative, Positive}
            \end{axis}
        \end{tikzpicture}
        \caption{Percentage of text chunks for negative and positive labels per continent.}
        \label{fig:distilbert_results_continent}
    \end{subfigure}
    
    \begin{subfigure}{\textwidth}
        \centering
        \begin{tikzpicture}
\begin{axis}[
    width=0.6\linewidth,
    ybar stacked,
    bar width=15mm,
    ymin=0,
        enlarge x limits=0.25,
    nodes near coords,
    legend style={at={(0.5,1.13)},
      anchor=north,legend columns=-1,draw=none},
    ylabel={\% of text chunks},
    symbolic x coords={Working, Non-working},
    xtick=data,
    ]
\addplot+[ybar,col1,fill=col1,text=black] plot coordinates {(Working,65) (Non-working,63)};
\addplot+[ybar,col4,fill=col4,text=black] plot coordinates {(Working,35) (Non-working,37)};
\legend{Positive,Negative}
\end{axis}
\end{tikzpicture}
        \caption{Percentage of text chunks for negative and positive labels per working and not working time hours.}
        \label{fig:distilbert_results_time}
    \end{subfigure}
    \caption{Results of the DistilBERT model for text chunk victims' responses per continent and working/non-working time hours.}
\end{figure}

To classify the users' sentiment into positive and negative, we used DistilBERT\footnote{\url{https://huggingface.co/distilbert-base-uncased-finetuned-sst-2-english}}, a transformer-based model without any further fine-tuning \cite{sanh2019distilbert}. Figure \ref{fig:distilbert_results} shows that most of the texts are classified as negative and are almost double the ones in the positive category. This indicates that victims have been negatively positioned towards the message of the phishing emails and Meta. According to Figure \ref{fig:distilbert_score}, it is also clear that there is a lot of confidence in the predictions of the model since, for almost all assessments, the score is close to one. We also investigated the users' demographics to drill down on these assessments. Figure \ref{fig:distilbert_results_continent} shows that the percentage of negative labels outnumbered the positive ones, with Oceania achieving the highest rate, while Africa has the lowest, reaching an absolute balance of positive and negative labels. Considering the time perspective, we split the timing into working and non-working hours. To this end, we considered an extended working timeframe of 08:00 a.m. to 18:00 p.m. Figure \ref{fig:distilbert_results_time} illustrates that the timing plays a little factor in the users' predisposition, as the difference is on the scale of 3\%, which can be marginally considered statistically significant for these percentages.

\begin{figure}[th]
    \centering
    \begin{subfigure}{\textwidth}
        \centering
        \begin{tikzpicture}
        \begin{axis}[
            ybar stacked,
            bar width=15mm,
            ymin=0,
            width=\textwidth, 
            height=0.5\textwidth,
            xlabel={Continents},
            ylabel={\% of text chunks},
            nodes near coords,
            symbolic x coords={Africa, America, Asia, Oceania, Europe},
            xtick=data,
            legend style={at={(0.5,1.15)}, 
              anchor=north,legend columns=-1,draw=none},
            ]
            \addplot[col1,fill=col1,text=black] coordinates {(Africa, 100) (America, 71) (Asia, 79) (Oceania, 82) (Europe, 73)};
            \addplot[col4,fill=col4,text=black] coordinates {(Africa, 0) (America, 29) (Asia, 21) (Oceania, 18) (Europe, 27)};
            \legend{Negative, Positive}
        \end{axis}
        \end{tikzpicture}
        
        \caption{Percentage of profanity text chunks for negative and positive labels per continent.}
        \label{fig:distilbert_profanity_results_continent}
    \end{subfigure}
  
    \begin{subfigure}{\textwidth}
        \centering
        \begin{tikzpicture}
\begin{axis}[
    width=0.6\linewidth,
    ybar stacked,
    bar width=15mm,
    ymin=0,
        enlarge x limits=0.25,
    nodes near coords,
    legend style={at={(0.5,1.13)},
      anchor=north,legend columns=-1,draw=none},
    ylabel={\% of text chunks},
    symbolic x coords={Working, Non-working},
    xtick=data,
    ]
\addplot+[ybar,col1,fill=col1,text=black] plot coordinates {(Working,76) (Non-working,67)};
\addplot+[ybar,col4,fill=col4,text=black] plot coordinates {(Working,24) (Non-working,33)};
\legend{Positive,Negative}
\end{axis}
\end{tikzpicture}
        \caption{Percentage of profanity text chunks for negative and positive labels per working and not working time hours.}
        \label{fig:distilbert_profanity_results_time}
    \end{subfigure}
    
    \caption{Statistics of the DistilBERT model for text chunk victims’ responses on the use of profanity.}
    \label{fig:profanity_comparison}
\end{figure}
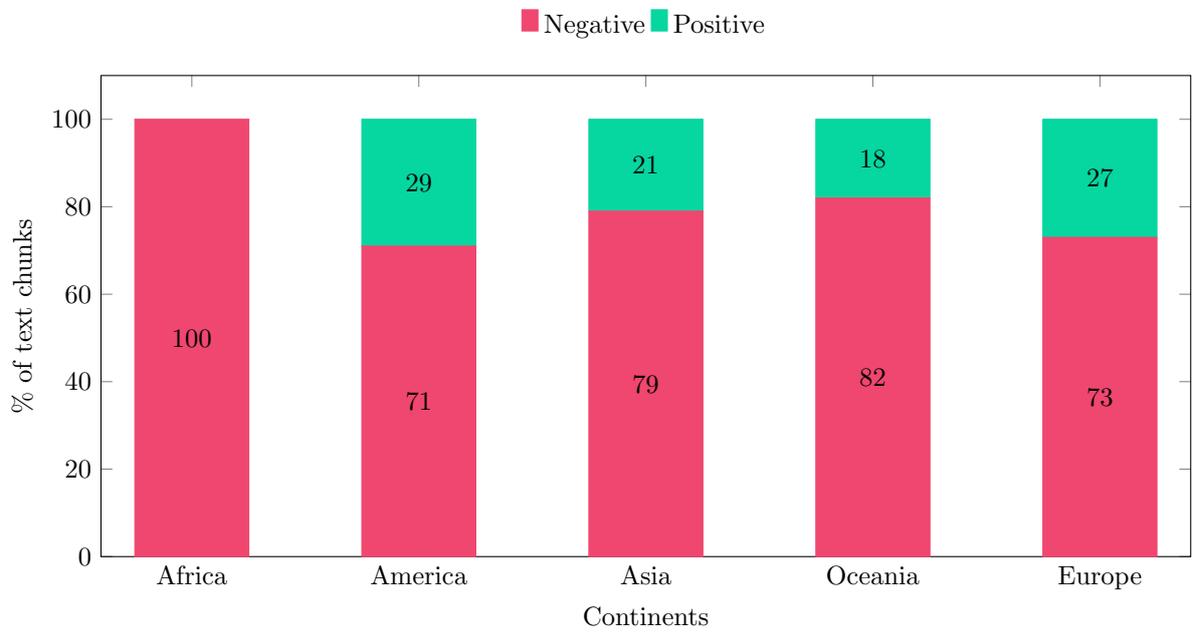
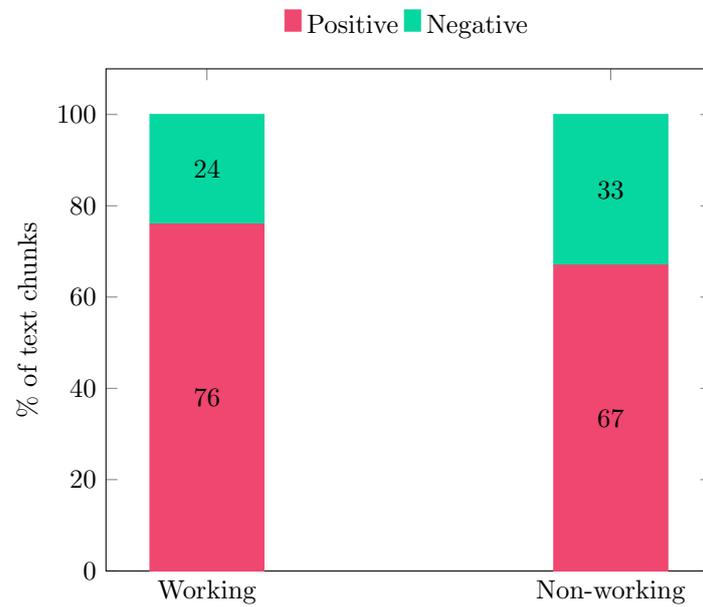

Regarding the victims' responses, which contain profanity, in Figure \ref{fig:distilbert_profanity_results_continent}, the distribution of negative and positive sentiments across the continents are illustrated. Notably, Africa is the continent with the highest percentage, reaching 100\% negative profanity responses, while the other continents have a high rate, which is more than 70\%. Europe has the lowest percentage, about 70\%. The timing factor was also examined in this case, showing that, as illustrated in Figure \ref{fig:distilbert_profanity_results_time}, the percentage of negative profanity texts outweighs the positive ones during working hours. We interpret the use of profanity along with negative sentiment as an indication that the users understood the scam and responded to the phishers accordingly, as validated with some random samples.

Examining the proportion of positive and negative sentiments across the week days, there was no noticeable variation, with the same applying to variations between working hours and beyond them. Similarly, the sentiment scores for both classes are relatively consistent across continents, with minor variations.

\subsection{Emotion analysis}
\begin{figure}[th!]
    \centering
    \begin{subfigure}{\textwidth}
        \includegraphics[width=\textwidth]{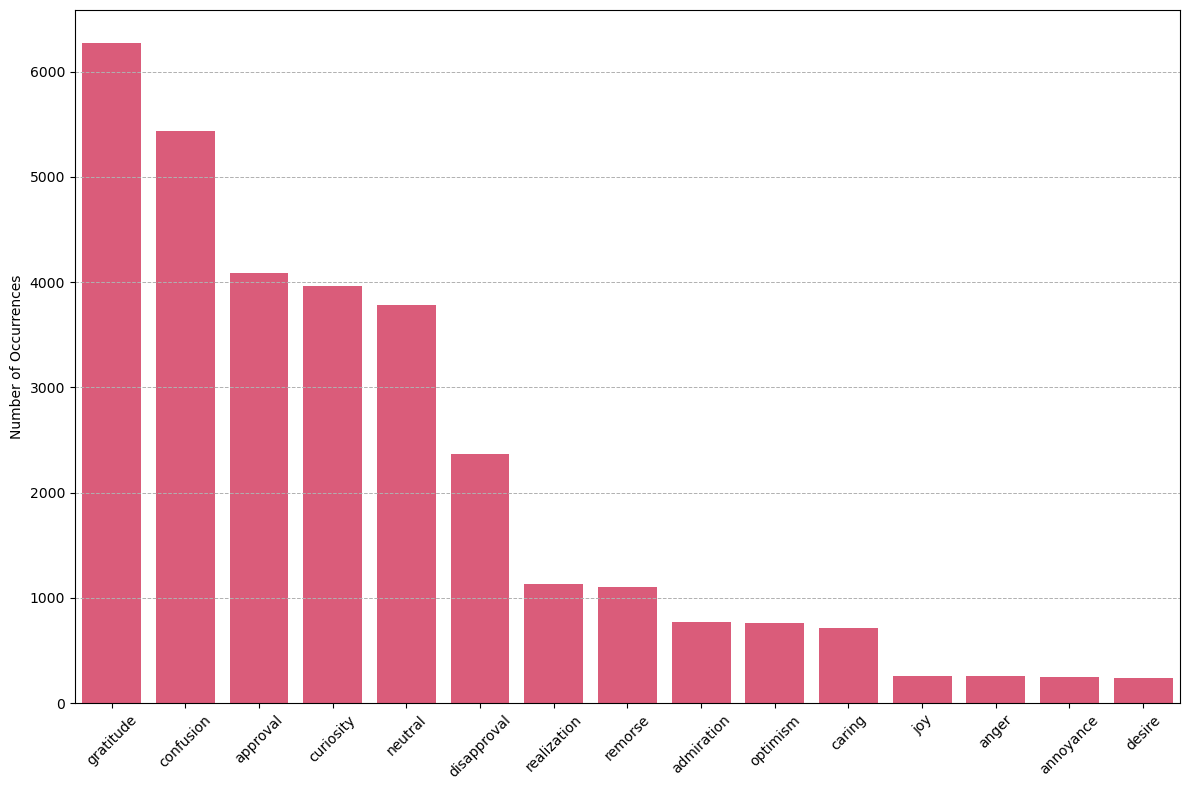}
        \caption{Distribution of text chunks per top-15 emotions.}
        \label{fig:emoroberta_results}
    \end{subfigure}
    \begin{subfigure}{\textwidth}
    \centering
        \includegraphics[width=.8\textwidth]{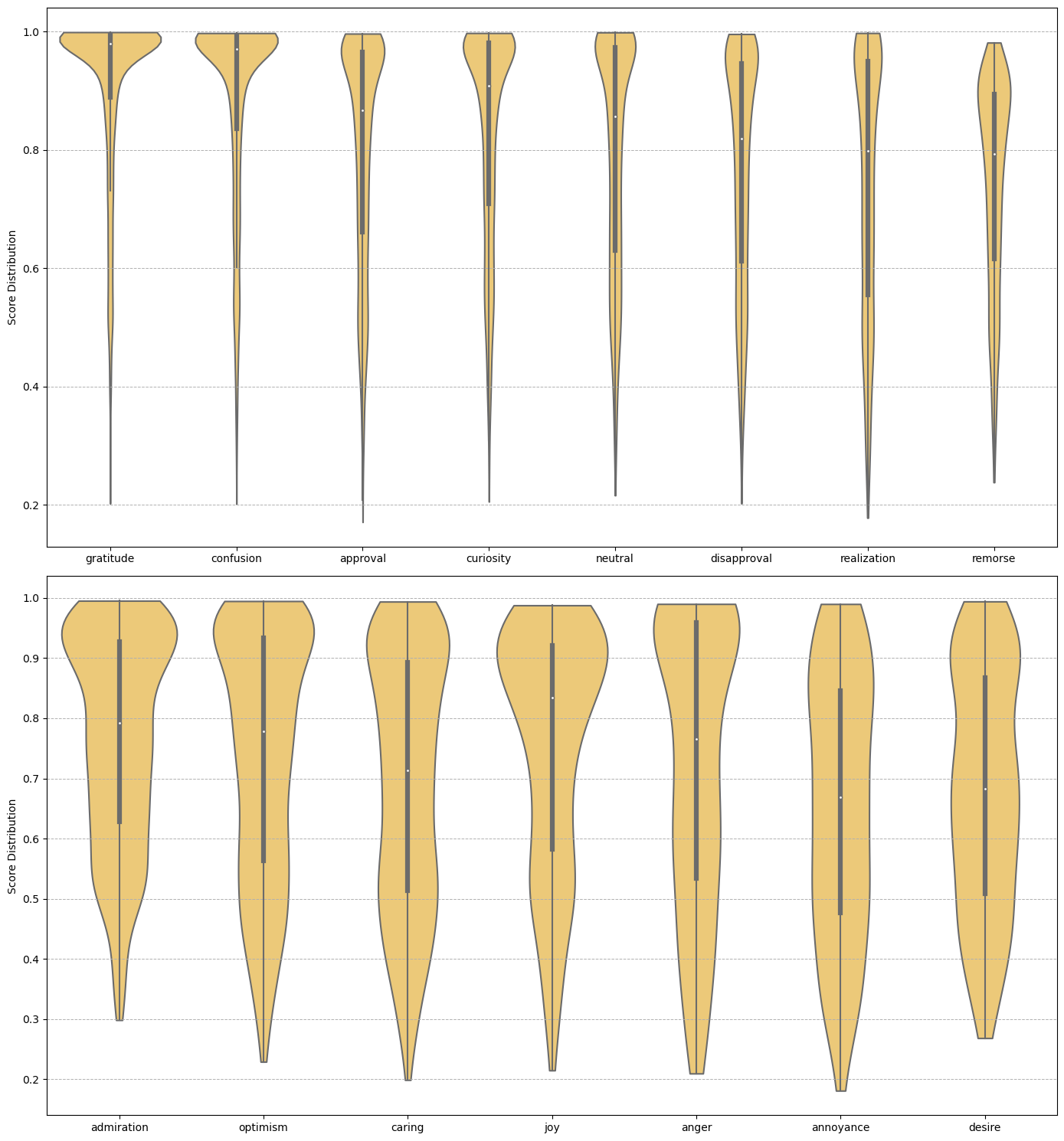}
        \caption{Distribution of scores for top-15 emotions .}
        \label{fig:emoroberta_score}
    \end{subfigure}
    
    \caption{Results of the EmoRoBERTa model for text chunk victims' responses.}
    \label{fig:emoroberta_combined}
\end{figure}

Another way to gain insight into phishing victims' behaviour and their emotional state is through emotion detection of their text inputs. To this end, we used the EmoRoBERTa\footnote{\url{https://huggingface.co/arpanghoshal/EmoRoBERTa}} transformer-based model without further fine-tuning it \cite{emoroberta}. This model classifies texts into 28 emotional categories, including but not limited to confusion, anger, and gratitude. Figure \ref{fig:emoroberta_results} shows the distribution of the analysed texts in these emotion categories, but for the sake of clarity, we kept only the top 15 measurements, as it is clear that for the rest the values are too low. Moreover, as observed in Figure \ref{fig:emoroberta_score}, it is clear that not all emotions have high confidence scores nor uniform distribution. It can be seen that the confidence scores of the top five emotions are close to one, while in other emotions, which are not so frequent in our dataset, such as relief, the score is low. Due to the fact that we have very little data for these emotions, we can say that they will not change the general picture of the dataset in the case of a wrong prediction.

Notably, the top four categories are gratitude, confusion, approval, and neutral. Taking into account the highest-ranking emotion categories, it seems that many victims, while confused, respect Meta's authority and seek to rectify Meta's decision to lock their accounts. In addition, they are grateful for the timely notification. The neutral emotion probably stems from the professional tone they try to have in their correspondence; see also further discussion below, and how they try to object to Meta's decision. We also examined possible correlations between emotions and demographics or timing. However, we did not observe any tangible differentiation. This indicates that demographic and timing factors did not affect the victims' sentiments.

\begin{figure}[th!]
    \includegraphics[width=\textwidth]{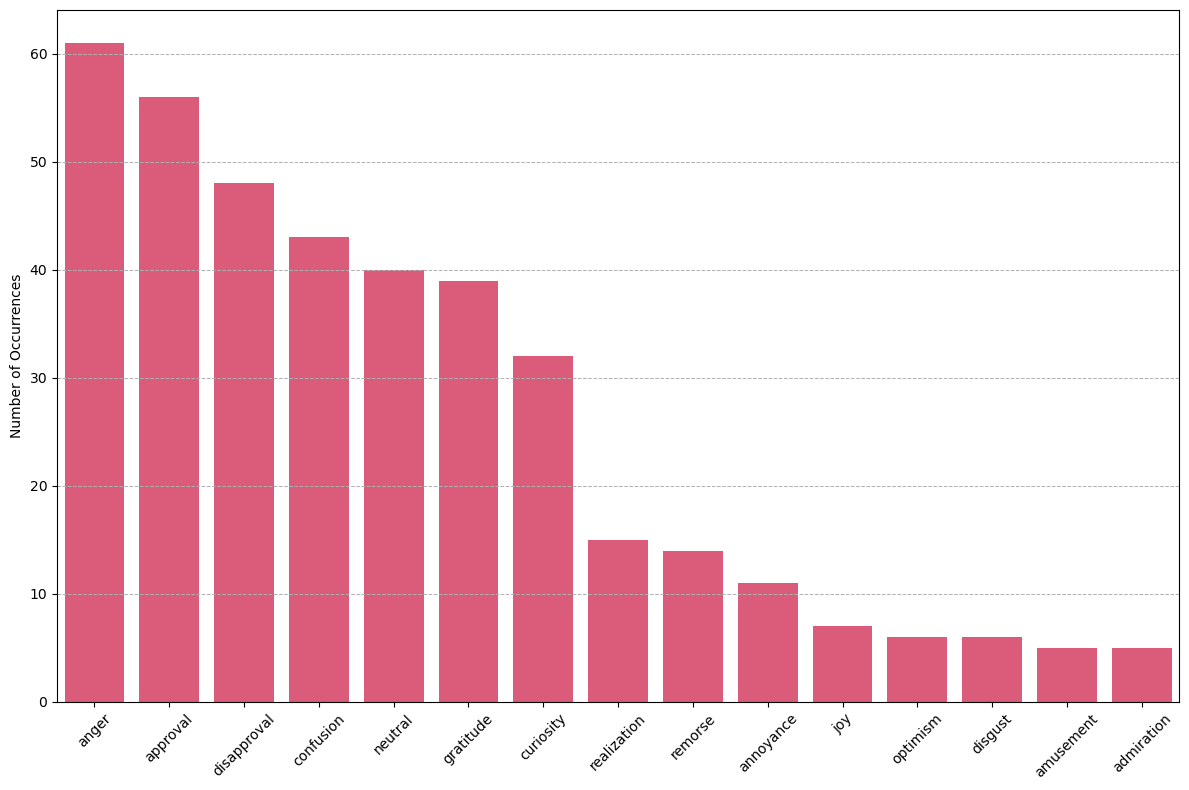} 
    \centering
    \caption{Distribution of profanity texts in top-15 emotions using the EmoRoBERTa model.}
    \label{fig:emoroberta_profanity}
\end{figure}

Regarding responses containing profanity, it can be seen in Figure \ref{fig:emoroberta_profanity} that most of them are categorised into the approval category, while other categories, such as disapproval, anger, curiosity, and confusion achieve a high rate. This means that many victims do not curse the alleged sender of the associated phishing email message, yet their responses contain profanity. For instance, the victims may have stated in their appeal that their account does not host nude pictures or sexual content, yet, the words \say{nude} and \say{sexual} are often filtered for profanity. The second highest-ranked category is anger, which indicates that several users perceived the email message that they received as a scam and responded angrily, using profane expressions. We found no tangible variations between emotions and continents, meaning the users' emotions were globally consistent. 

\subsection{Tone analysis}
Given the exceptional results of large language models (LLMs) in various tasks compared to humanly produced results \cite{gilardi2023chatgpt,patsakis2023man}, we used OpenAI's GPT 4 API to classify the tone of the provided texts into the following categories: polite, impolite, formal, informal, sarcastic, sad, optimistic, pessimistic, urgent, calm, angry, affectionate, apathetic, defensive, and aggressive. 

\begin{figure}[th]
    \centering
    \includegraphics[width=\textwidth]{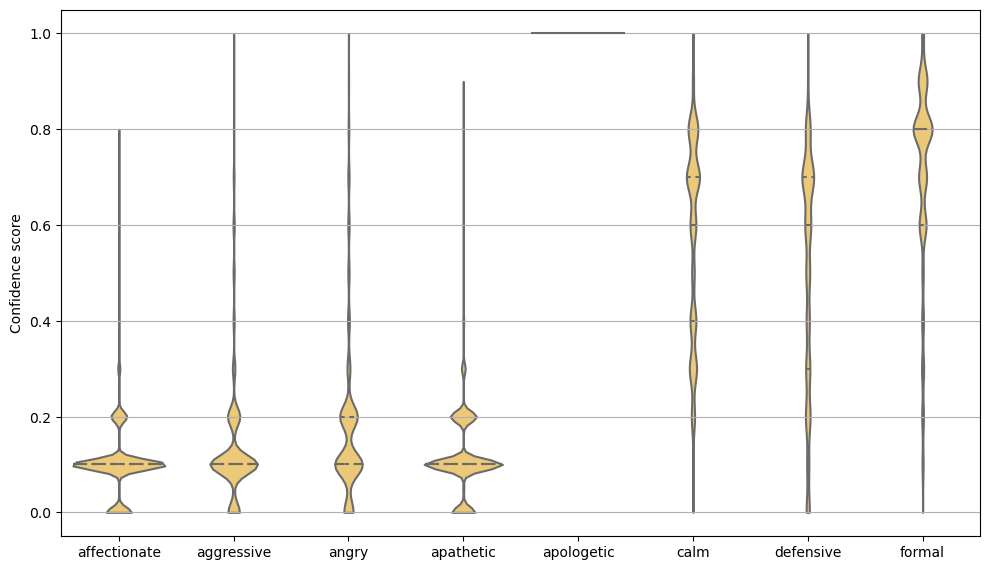}
    \includegraphics[width=\textwidth]{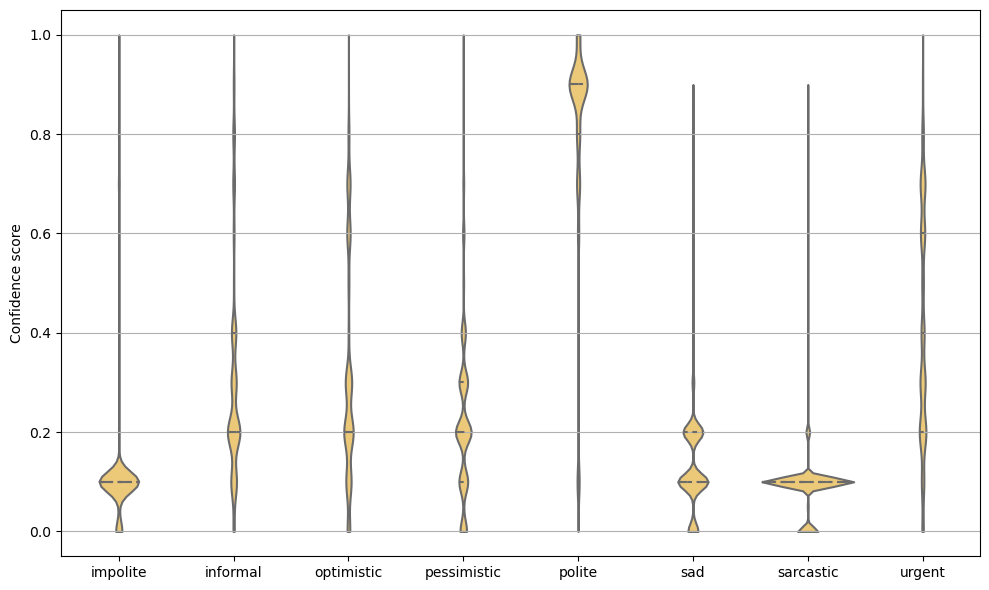}
    \caption{Tonality classification of anonymised texts with GPT-4.}
    \label{fig:tonality}
\end{figure}

Regardless of OpenAI's claimed policies\footnote{\url{https://openai.com/policies/usage-policies}}, since some users had provided personal information in their texts, we processed the texts with Textwash \cite{kleinberg2022textwash} to replace any personally identifiable information before submitting them to GPT 4. Thus, we anonymised the texts and submitted them to GPT 4 with the task illustrated in Table \ref{tbl:gpt_task}, requesting the confidence level for each classification and using a low temperature (0.2) to minimise the possible randomness of the output. The results are illustrated in Figure \ref{fig:tonality} as violin plots to illustrate the distribution of the confidence level per tone. As observed, the users' input is almost always apologetic and polite, rather formal, calm, and defensive, while users often express their urgency in their appeal. The latter indicates that the vast majority of users, regardless of whether they provided the correct input, feel that they are indeed appealing to Meta and try to appear professional in their requests, apologising for something that they obviously have not done. Clearly, angry, aggressive, and impolite users are a very small minority. Notably, most of them are users who have understood the scam and react to phishers. Therefore, we can safely say that the relevant campaigns were very convincing and engaging.

\begin{table}[th]

\centering
\begin{tabular}{|p{\textwidth}|}
    \hline
    \cellcolor{col2!10}\textbf{System:} You are a text analyser. Classify the text as polite, impolite, formal, informal, sarcastic, sad, optimistic, pessimistic, urgent, calm, angry, affectionate, apathetic, defensive, and aggressive. Return only the output as a Python dictionary with the confidence level for each classification. Do not justify your output.\\
    \hline \textbf{User:} I don't know why this is. There have been no such announcements before. I haven't changed anything when I advertise.\\   
    \hline \textbf{Assistant:} \begin{lstlisting}[language=json]
{"polite": 0.75, "impolite": 0.1, "formal": 0.8,
  "informal": 0.2, "sarcastic": 0.1, "sad": 0.3,
  "optimistic": 0.1, "pessimistic": 0.6, "urgent": 0.2,
  "calm": 0.8, "angry": 0.2, "affectionate": 0.05,
  "apathetic": 0.2, "defensive": 0.7, "aggressive": 0.1}
\end{lstlisting}\\
    \hline
\end{tabular}
\caption{An example of a categorisation response produced by GPT.}
\label{tbl:gpt_task}
\end{table}

\section{Operational issues}
\label{sec:ops}
Further to the operational security issues that enabled us to collect the data, we noticed many other operational issues. For instance, the phishing form did not have a proper input validation mechanism. As a result, one could easily inject arbitrary input to almost any field. Indeed, as illustrated in Table \ref{tbl:victim_stats}, one user did not actually use a browser but a command line tool (curl) to inject several hundreds of rows using curl, swearing at the phishers. Moreover, several rows contain obviously inaccurate data, e.g., non-numeric TFAs.

Additionally, several rows indicate that they are tests from the phishers to determine that their deployment is working. Indicatively, some rows originate from local IPs, which, of course, can only be performed by someone who has local access to the machine. Moreover, the origin of other IPs of the testing rows narrows down the phishers' origin to two countries. More precisely, there seem to be two users controlling the platform, one from a Balcan and one from an ASEAN country.

Finally, in the phishing sites, apart from a design which points to Meta in terms of user interface, the user experience and the content do not meet the Meta standards. For instance, the links and menus were not operational, something that would not be expected from a real Meta web page, let alone one designed for an important service or user notification. Therefore, page visitors should have immediately understood that something was wrong and leave.

\section{Conclusions}
Phishing attacks are continuously increasing and constitute a major threat for modern organisations. Merely depending on technical means to counter them is not enough, as this ends up being a number game. Even if researchers and companies claim the precision and accuracy of their methods to be greater than 99\%, the remaining percentage, when projected to the sheer volume of phishing emails that are distributed every day, means that several thousands of emails are expected to bypass these filters. Since this should be considered a de facto truth, it is up to organisations to properly train their personnel appropriately and mitigate this risk. 

Therefore, understanding the human factor in such events is crucial in properly addressing this threat. Our research, beyond merely presenting a successful phishing attack, studies the different perspectives, specifically focusing on human factors. To the best of our knowledge, this is the first academic work that does this kind of research in a real-world case and on such a scale. Our findings reveal several demographic patterns showing specific trends subject to the residence of the victims, e.g., what time the victim would respond to a phishing email. Moreover, we observe that timing is a huge factor, as victims would most likely respond to phishing emails during working hours and are most likely to do it in the first days of the week. The latter implies that users are most likely stressed to respond to emails stacked in their mailbox during the weekend, augmenting the probability of being exploited in this way. 

Nevertheless, other aspects of the phishing campaigns also reveal human traits. For instance, our research reveals that phishing victims do not follow the best password practices. Beyond falling for the phishers' bait, many of them not only use weak passwords, but more than half use passwords already leaked more than two years ago. Practically, this means that it is very likely that they have been repeatedly scammed. In fact, this persistence is exhibited by the fact that a significant part of the victims would interact with the phishing "platform" more than once, even in the scope of another phishing email of the same campaign. Thus, we can safely argue that a significant number of phishing victims tend to re-victimisation due to their poor choices in cyber hygiene.

Of specific interest is the fact that the victims did not rationalise the content of the phishing campaign that targeted them. One should at least consider that the involvement of all three organisations does not make much sense, especially since one of them (Meta), has never used the others to offer them other services. The users should have understood that Meta has no good reason to use Salesforce to send them an email, as they already receive emails directly from Meta for other notifications. Similarly, there is no good reason why Meta would use Google's platform to host its content. Even more, by simply logging into their Meta account, they would easily check that there would be no notification to warn them about an upcoming service termination. Finally, a service provider would not need to obtain the users' passwords through a public form. This illustrates that while there were many red flags, thousands of people fell for the bait, signifying that any performed awareness campaigns should be more targeted and practical to make users even more robust or at least more sceptical towards phishing campaigns.

The above portrays the victims as people who, through carelessness or anxiety due to work pressure, provided their credentials and sensitive information. The way they respond to the campaign signifies that the users did not realise at any point that they were being scammed, but on the contrary, despite all other visual and logical evidence, they proceeded with providing the sensitive information. It is notable that the demographics played a minor role in the emotion of the users; however, the common denominator seems to be a general lack of cybersecurity hygiene, as most manifested by the poor password choices and the persistence in responding to other emails of the campaign or returning to interact with the platform. Hence, it is fair to argue that phishing awareness campaigns should be more focused and try to help users who show other bad cybersecurity practices to change their posture.

\section*{Acknowledgements}
This work was supported by the European Commission under the Horizon Europe Programme, as part of the projects LAZARUS (Grant Agreement no. 101070303) and HEROES (Grant Agreement no. 101021801), and under the ISF-P Programme, as part of the projects CTC (\url{https://ctc-project.eu/}) (Grant Agreement no. 101036276). This work was also supported by the COST Action GoodBrother, Network on Privacy-Aware Audio- and Video-Based Applications for Active and Assisted Living (CA 19121). 

The content of this article does not reflect the official opinion of the European Union. Responsibility for the information and views expressed therein lies entirely with the authors.

\end{document}